\documentclass[reprint,notitlepage,superscriptaddress,nobibnotes,amsmath,amssymb,aps,prd,noeprint,floatfix]{revtex4-2}

\usepackage{easyReview}
\usepackage{graphicx}
\usepackage{dcolumn}
\usepackage{bm}
\usepackage{mathtools}
\usepackage[utf8]{inputenc} 
\usepackage{xcolor}
\usepackage{lineno}
\usepackage{aas_macros}
\usepackage{url}
\usepackage[colorlinks]{hyperref}
\usepackage{slashed}
\usepackage{calc}
\usepackage{amsmath}
\usepackage{amssymb}
\usepackage[most]{tcolorbox}
\definecolor{dodgerblue}{HTML}{1E90FF}
\definecolor{ferrarired}{HTML}{ff2800}
\definecolor{olive}{HTML}{808000}
\definecolor{maroon}{HTML}{800000}
\newcommand\orcidlink[1]{\href{https://orcid.org/#1}{\includegraphics[scale=0.006]{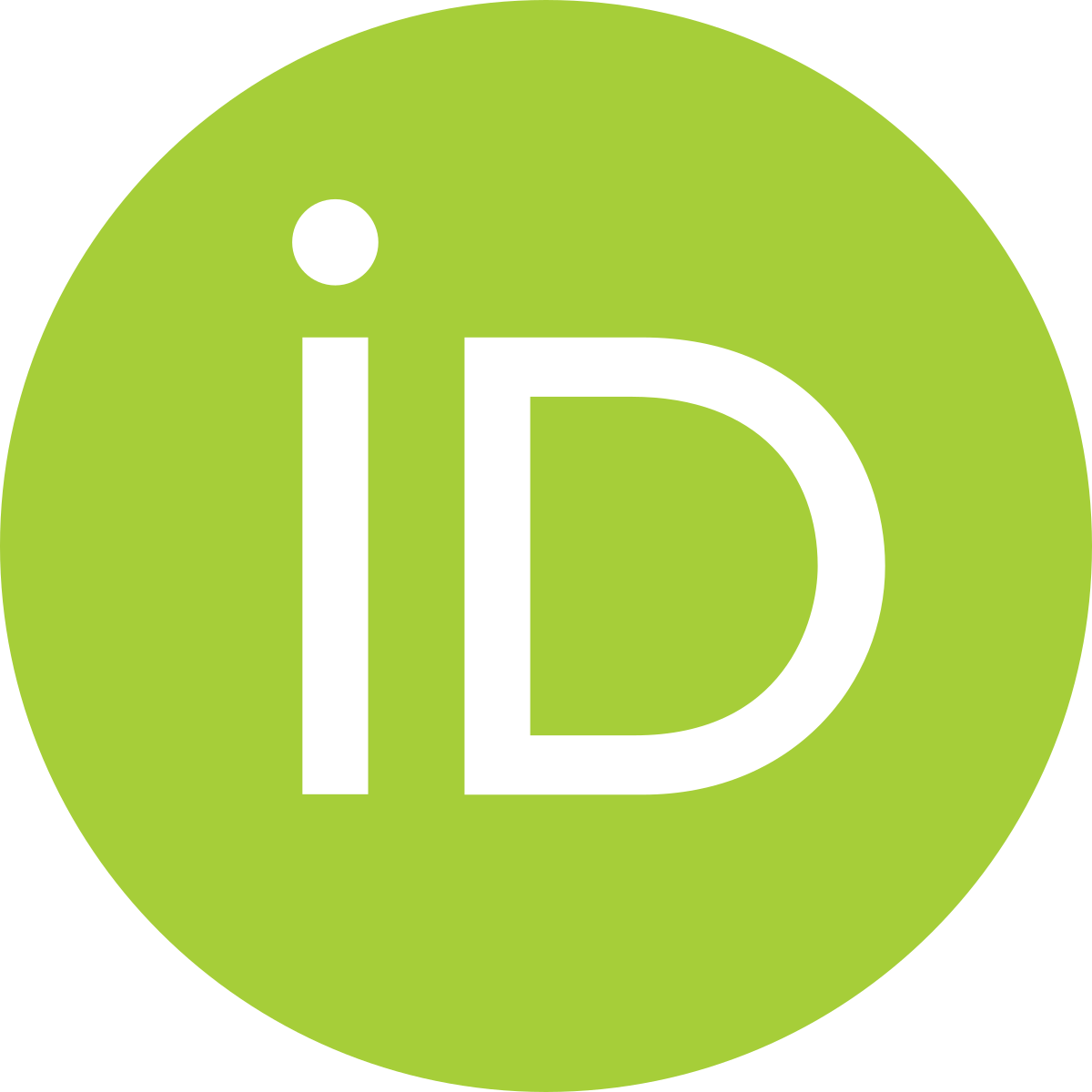}}}

\makeatletter
\newcommand{\distas}[1]{\mathbin{\overset{#1}{\kern\z@=}}}
\makeatother

\makeatletter
\def\namedlabel#1#2{\begingroup
	\def\@currentlabel{#2}%
	\label{#1}\endgroup}
\makeatother

\hypersetup{
     colorlinks=true,
     linkcolor=maroon,
     filecolor=maroon,
     citecolor = maroon,      
     urlcolor=maroon,
     }

\begin{document}

\title{Detecting non-Gaussian gravitational wave backgrounds: a unified framework}

\newcommand{\infnpisa}{\affiliation{INFN Sez.~Pisa, Largo B. Pontecorvo 3, I-56127 Pisa, Italy}}

\newcommand{\unipi}{\affiliation{Universit\`{a} di Pisa, Dipartimento di Fisica ``E. Fermi'', Largo B. Pontecorvo 3,  I-56127 Pisa, Italy}}
\newcommand{\bham}{\affiliation{Institute for Gravitational Wave Astronomy \& School of Physics and Astronomy, University of Birmingham, Birmingham, B15 2TT, UK}}
\newcommand{\milan}{\affiliation{Dipartimento di Fisica ``G. Occhialini'', Universit\'a degli Studi di Milano-Bicocca, Piazza della Scienza 3, 20126 Milano, Italy} \affiliation{INFN, Sezione di Milano-Bicocca, Piazza della Scienza 3, 20126 Milano, Italy}}
\author{Riccardo Buscicchio~\orcidlink{0000-0002-7387-6754}}
\email{riccardo.buscicchio@unimib.it}
\milan
\bham

\author{Anirban Ain~\orcidlink{0000-0003-4534-4619}}
\infnpisa
\unipi

\author{Matteo Ballelli~\orcidlink{0000-0003-1512-5423}}
\infnpisa
\unipi

\author{Giancarlo Cella~\orcidlink{0000-0002-0752-0338}}
\infnpisa

\author{Barbara Patricelli~\orcidlink{0000-0001-6709-0969}}
\infnpisa
\unipi

\date{\today}
\begin{abstract}
	We describe a novel approach to the detection and parameter estimation of a non\textendash Gaussian stochastic background of gravitational waves. 
	The method is based on the determination of relevant statistical parameters using importance sampling. 
	We show that it is possible to improve the Gaussian detection statistics, by simulating realizations of the expected signal for a given model.
	While computationally expensive, our method improves the detection performance, leveraging the prior knowledge on the expected signal, and can be used in a natural way to extract physical information about the background.
	We present the basic principles of our approach, characterize the detection statistic performances in a simplified context and discuss possible applications to the detection of some astrophysical foregrounds. 
	We argue that the proposed approach, complementarily to the ones available in literature might be used to detect suitable astrophysical foregrounds by currently operating and future gravitational wave detectors.
\end{abstract}

\maketitle

\section{Introduction}
\label{sec:Introduction}
Over the past seven years, the Advanced Laser Interferometer Gravitational-Wave Observatory (LIGO)~\cite{2015CQGra..32g4001L} and Advanced Virgo~\cite{2015CQGra..32b4001A} have collected data and released, together with the KAGRA collaboration~\cite{2019NatAs...3...35K}, three incremental catalogues of gravitational-wave (GW) detections, amounting to a total of 90 confident events ~\cite{2021arXiv211103606T}.
In addition, the LIGO, Virgo and KAGRA collaboration (LVKC) have performed a population study on a subset of 76 of them~\cite{2021arXiv211103634T}.
Further upgraded second generation interferometers~\cite{collaboration2013prospects} and third generation GW detectors, such as the Einstein Telescope (ET,~\cite{2010CQGra..27s4002P}), will possibly become operational during the next decade: 
these  experiments promise to be sensitive enough to observe both the cosmological and astrophysical stochastic gravitational wave background (SGWB).
In addition, when the Large Interferometer Space Antenna (LISA)~\cite{2017arXiv170200786A} will be operational, our sensitivity to astrophysical GW transients will broaden to lower frequencies and new source categories.

The superposition from various unresolved astrophysical and cosmological sources generates a SGWB.
Searches for such a stochastic background have been performed on available data:
no evidence for a SGWB has been found; nonetheless upper limits on its cosmological energy density have been placed~\cite{2021PhRvD.104b2004A,2021arXiv211103634T}. 
Among the sources which may contribute to the SGWB: core-collapse supernovae~\cite{2005PhRvD..72h4001B,2004MNRAS.351.1237H,2006PhRvD..73j4024S}; 
neutron stars~\cite{1999MNRAS.303..258F,2001A&A...376..381R,2013PhRvD..87f3004L}; 
compact binary coalescences~\cite{2013MNRAS.431..882Z,2011PhRvD..84l4037M,2012PhRvD..85j4024W,2011PhRvD..84h4004R}; 
binary white dwarfs~\cite{2003MNRAS.346.1197F};
cosmic strings~\cite{1992PhRvD..45.3447C,2005PhRvD..71f3510D,2010PhRvD..81j4028O} and gravitational waves produced during inflation~\cite{1993PhRvD..48.3513G,2007PhRvL..99v1301E,2012PhRvD..85b3534C} or by primordial black holes~\cite{2018arXiv181211011C}.
A detection of the cosmological SGWB would give very important constraints on the earliest epochs  of the Universe, while the detection of an astrophysical SGWB would provide key information about the sources generating it, e.g. the merger rate of compact binary systems, the star formation history~\cite{2020ApJ...896L..32C, 2022A&A...660A..26B} or the occurrence of gravitational-wave lensing~\cite{2021ApJ...923...14A, PhysRevLett.125.141102}.

Searches for SGWBs typically assume that the background is Gaussian, based on the central limit theorem (see, e.g.,~\cite{TheLIGOScientific:2016dpb,LIGOScientific:2019vic}).
However, if the rate of events generating the background is not sufficiently high compared to their duration or frequency bandwidth, a non-Gaussian background is expected, characterized by discontinuous or intermittent signals.
For instance, predictions based on population modelling suggest that, for many realistic astrophysical models, there may not be enough overlapping sources, resulting in the formation of such a non-Gaussian background (see, e.g.,~\cite{2011ApJ...739...86Z, 2020MNRAS.491.4690M}).
Furthermore, it has been shown that the background from cosmic strings could be dominated by a non-Gaussian contribution arising from the closest sources~\cite{2005PhRvD..71f3510D}. 

In the past decades several methods to search for non-Gaussian SGWB have been proposed.
For instance, the authors of Ref.~\cite{Drasco:2002yd} derived an algorithm suitable for the detection of a non-Gaussian component in a SGWB observed by two co-located and co-aligned detectors with white Gaussian noise.
Later,the author of Ref.~\cite{Thrane:2013kb} introduced a maximum likelihood estimator to be used in a more realistic case of a network of spatially separated interferometers with coloured, non-Gaussian noise.
Most recently, the authors of Ref~\cite{Smith:2018PhRvX...8b1019S} devised a Bayesian search strategy for a background of unresolved binaries.
Other approaches have also been explored (see, e.g.,Refs.~\cite{Lionel:2015kpa, Regimbau:2011bm,Martellini:2014xia}), constructing alternative parametrizations for SGWB non-Gaussianities.
In the context of LISA, various pipelines for the detection and characterization of an astrophysical SGWB have been developed~\cite{2021JCAP...01..059F, 2022arXiv220407349G, PhysRevD.104.043019}, parametrizing a certain level on non-Gaussianities in the signal model. The expected level has also been assessed for confusion noise arising from extreme mass ratio inspirals (EMRIs) and Galactic binary white dwarfs~\cite{Racine:2007gv}.

In this paper, we explore a novel approach for a detection of non-Gaussian SGWBs --inherently complementary to the ones available in literature~\cite{Smith:2018PhRvX...8b1019S}-- using a detailed stochastic model of the underlying signal population.
The paper is organized as follows: in Sec.~\ref{sec:The-statistical-problem} we discuss the basic principles for the detection of a SGWB, and we give examples of application for the case of an isotropic background;
after a discussion of the Neyman--Pearson detection statistic (DS) in a frequentist context (Sec.~\ref{subsec:NPdetector}) we show how a Bayesian analysis of a non-Gaussian stochastic background can be implemented (Sec.~\ref{subsec:Bayesian-approach});
in Sec.~\ref{sec:toymodel} we discuss a simplified model for a non-Gaussian stochastic background, with the purpose of estimating the improvement in detection performance of the proposed approach;
in Sec.~\ref{sec:AstroSB} we give details about the application to a more realistic case, namely an isotropic stochastic background of astrophysical origin; we show how this can be represented by a (generalized) point process (Sec.~\ref{subsec:point-process}), and give details about the stochastic sampling procedure required by the inference method (Sec.~\ref{subsec:Importance-sampling});
finally, in Sec.~\ref{sec:Conclusions-and-perspectives} we draw some conclusions pointing at possible future developments, in particular toward applications to the nonisotropic case;
in Appendix~\ref{app:supplemental} we provide detailed proofs of results shown in the main text. Some are available in literature (see e.g. Ref.\cite{2022Galax..10...34R} and references therein), nonetheless we choose to reproduce them to ensure consistency of notation across the text.

\section{The statistical problem}
\label{sec:The-statistical-problem}

In this paper, observations are written as the sum of signal and noise, however in our case it is more convenient to write the data collected by a network of detectors in a slightly different form, namely
\begin{equation}\label{eq:decomposition}
	s_{i}^{{\cal A}}=g_{i}^{{\cal A}}+h_{i}^{{\cal A}}+n_{i}^{{\cal A}} ,
\end{equation}
where $g$ is a Gaussian part of the stochastic signal and $h$ a non-Gaussian one, while $n$ is the noise of the detectors. We assume statistical indipendence among the three components $g$, $h$ and $n$. We assume also that they have zero mean. A nonzero average is observationally irrelevant and can be removed in the time domain. In the frequency domain it could model a spurious non--stochastic contamination that should be removed before the analysis.
	
In this paper we always assume an additive and Gaussian noise, although an extension is possible to account for transient non-stationarities arising from the noise.
Capital indices label the detector while lowercase ones enumerate generically the data series: 
we will specialize it if needed by explicitly writing our expressions in time or frequency domain.
It is worth emphasizing that the decomposition in Eq.~\eqref{eq:decomposition} is not unique: one can always add and subtract an arbitrary Gaussian contribution to $g$ and from $h$. This is a feature arising from the inherent modeling freedom for $h$ in Eq.\eqref{eq:decomposition} and it is not related to the noise properties.

Under our hypotheses the noise is described by a multivariate Gaussian probability distribution that we can write as
\begin{equation}
	\label{eq:pnoise}
	p_{n}\left[n_{i}^{{\cal A}}\right]  = {\cal N}_n \exp\left(-\frac{1}{2}{\cal W}_n(n,n)\right),
\end{equation}
where ${\cal N}_n$ is a normalization constant and for future convenience we defined the scalar product over detectors and data indices
\begin{align}
	{\cal W}_x(u,v)&\equiv \sum_{{\cal A},{\cal B}}\sum_{i,j}\left[\mathbb{C}_{xx}^{-1}\right]_{ij}^{{\cal A}{\cal B}}u_{i}^{{\cal A}}v_{j}^{{\cal B}}, \\ 
	\left[\mathbb{C}_{xy}\right]_{ij}^{{\cal AB}} &\equiv \left\langle x_{i}^{{\cal A}}y_{j}^{{\cal B}}\right\rangle .\label{eq:defC}	
\end{align}
Hereafter, following Einstein's convention on repeated indices, we drop the summation symbols over data and detectors indices.
$\mathbb{C}_{nn}$ is the noise cross\textendash correlation array, so $\mathcal{W}_n$ in Eq.~\eqref{eq:pnoise} is the Wiener match between $u$ and $v$ with respect to the noise $n$~\cite{jaranowski2009analysis}. 
Explicitly we can write
\begin{align}
	{\cal N}_x & = \exp\left(-\frac{1}{2}\text{Tr}\ln2 \pi\mathbb{C}_{x}\right),\\
	\left[{\mathbb{C}}_{xy}\right]_{ij}^{{\cal AB}}&\approx\left[\check{\mathbb{C}}_{xy}\right]_{ij}^{{\cal AB}}  = \overline{x_{i}^{{\cal A}}y_{j}^{{\cal B}}},\label{eq:estimator}
\end{align}
where the trace is performed over detectors and data indices.
For simplicity, in autocorrelations $\mathbb{C}_{xx}$ we drop a redundant index, therefore denoting them $\mathbb{C}_{x}$.

In Eq.~\eqref{eq:estimator} and in what follows we often replace the true cross\textendash correlations $\mathbb{C}_{xy}$ \textendash~a theoretical expectation value defined through the model and frequently unmeasurable \textendash~with estimators obtained from the data. We label them with an overhead check. Correspondingly, averaging over the data indices is denoted with an overline.
\begin{equation}
	\left[ \check{\mathbb{C}}_{xy} \right]_{ij}^{{\cal AB}} = \overline{x_{i}^{{\cal A}}y_{j}^{{\cal B}}}.
\end{equation}
Because of statistical fluctuations, the uncertainty on $\check{\mathbb{C}}_{xy}$ can be improved by averaging over chunks of data: as we show in Eq.~\eqref{eq:prob_detection}, this comes at the cost of a reduced probability of detection.

We model the stochastic signal described by $g$ as Gaussian with a probability distribution analogous to Eq.~\eqref{eq:pnoise}, namely
\begin{equation}
	\label{eq:psigg}
	p_{g}\left[g_{i}^{{\cal A}}\right] = {\cal N}_g \exp\left(-\frac{1}{2}{\cal W}_g(g,g)\right).
\end{equation}
However, we make no statistical hypothesis about the remainder $h$, which can be described by a generic probability distribution $p_{h}\left[h_{i}^{{\cal A}}\right]$.
Then we can write the probability distribution for the observed signal as a convolution between $p_{n}$, $p_{g}$ and $p_{h}$, namely
\begin{multline}\label{eq:start}
	p_{s}\left[s\right]\!=\int_h\int_g p_{s}\left[s \mid h,g\right]p_{h}\left[h\right]p_{g}\left[g\right] \\ 
	={\cal N}_n{\cal N}_g\int_h\int_g \!p_{h}\!\left[h\right]\!e^{-\frac{1}{2}{\cal W}_n(s-h-g,s-h-g)-\frac{1}{2}{\cal W}_g(g,g)}.
\end{multline}
The Gaussian integral over $g$ can be performed explicitly. By virtue of Woodbury's identity (see Appendix \ref{app:supplemental} for details), we observe that
\begin{equation}\label{eq:wiener}
	{\cal W}_{n+g}(u,v) = {\cal W}_{n}(u,v) - {\cal G}(u,v)
\end{equation}
or equivalently
\begin{align}\label{eq:WoodIdent}
	\mathbb{C}_{n}^{-1}-\mathbb{C}_{n}^{-1}\left(\mathbb{C}_{g}^{-1}+\mathbb{C}_{n}^{-1}\right)^{-1} \mathbb{C}_{n}^{-1}=\mathbb{C}_{n+g}^{-1},
\end{align}
where we have defined for future convenience
\begin{align}
	{\cal G}(u,v) & \equiv \mathbb{G}^{\cal A \cal B}_{ij} u_i^{\cal A} v_j^{\cal B},\\
	\label{eq:defG}
	\mathbb{G} & \equiv \mathbb{C}_n^{-1} \left( \mathbb{C}_n^{-1}+\mathbb{C}_g^{-1}  \right)^{-1} \mathbb{C}_n^{-1},
\end{align}
which go to zero when $g=0$.
Using Eq.~\eqref{eq:WoodIdent} the integral further simplifies to
\begin{equation}\label{eq:Gaussian_integral}
	p_{s}\left[s\right]  =  {\cal N}_{n+g} \int_h p_{h}\left[h\right] e^{-\frac{1}{2}{\cal W}_{n+g}(s-h,s-h)}.
\end{equation}

The key point is that we can rewrite $p_{s}$ as
\begin{equation}
	p_{s}\left[s\right]={\cal N}_{n+g}  \left\langle e^{-\frac{1}{2}{\cal W}_{n+g}(s-h,s-h)}\right\rangle, \label{eq:SignalProbability}
\end{equation}
where the expectation value $\left\langle \cdots\right\rangle$ is evaluated over an ensemble of realizations for the non-Gaussian part $h$ of the SGWB.
Note that this expectation value is evaluated at fixed data $s$, which is considered here an independent variable.

While it is difficult to write an explicit expression for $p_{h}$ in the non\textendash Gaussian case, realizations of a stochastic background $h$ can be simulated. 
This opens up the possibility of evaluating $p_{s}$ and connected quantities related to DS and parameter estimation procedures.

\subsection{\label{subsec:NPdetector}The Frequentist approach}
As a first example we show an expression for the optimal Neyman\textendash Pearson DS~\cite{1933RSPTA.231..289N}, under the hypothesis of a known background and a known noise. The two hypotheses to be tested are
\begin{description}
	\item [{${\cal H}_{1}$}] presence of a known stochastic background, with a given Gaussian part $g$ and a given non\textendash Gaussian one $h$.
	\item [{${\cal H}_{0}$}] absence of the background, $g=h=0$, which means $s=n$.
\end{description}
The DS is defined by the test statistic $\hat{Y}(s)>\lambda$ where
\begin{align}
	\hat{Y}(s) & \equiv\log\frac{p_{s}\left[s|{\cal H}_{1}\right]}{p_{s}\left[s|{\cal H}_{0}\right]} -\log \frac{{\cal N}_{n+g}}{{\cal N}_n}\! \label{eq:Ystat} \\
	& = \frac{1}{2} {\cal G}(s,s) + \log\left\langle e^{-\frac{1}{2}{\cal W}_{n+g}(h,h)}e^{{\cal W}_{n+g}(s,h)}\right\rangle .\label{eq:Ystatsimple}
\end{align}
We subtracted from the standard definition of $\hat{Y}(s)$ a data independent constant, whose effect can be compensated by a redefinition of the relation between the threshold $ \lambda $ and the false alarm probability~\cite{kay1993fundamentals}.
Note that the average in Eq.~\eqref{eq:Ystatsimple} is evaluated under the ${\cal H}_1$ hypothesis.

\subsubsection{Gaussian case\label{subsubsec:Gaussian}}

We discuss shortly the particular case of a Gaussian background, as this clarifies some aspects relevant in the following sections.
If the background is gaussian we can assume, without loss of generality, that $h=0$  and the optimal statistic is given by 
\begin{align}
	\label{eq:Ygauss}
	\hat{Y}(s)=&\frac{1}{2}{\cal G}(s,s) \nonumber \\
	\simeq&
	\frac{1}{2} \left[
	\mathbb{C}_{n}^{-1}\mathbb{C}_{g}\mathbb{C}_{n}^{-1}
	\right]_{ij}^{{\cal AB}}
	\!
	s_i^{\cal A}s_j^{\cal B}
	+\mathcal{O}(\|\mathbb{C}_g\mathbb{C}_n^{-1}\|^2),
\end{align}
where we expanded Eq.~\eqref{eq:defG} to lowest order, under the hypothesis that the SGWB power spectrum is much smaller than every detector's noise spectrum. 
While the frequentist approach makes direct use of such assumption, the corresponding Bayesian approach in Sec.~\ref{subsec:Bayesian-approach} does not assume it, hence making it suitable in other contexts.
As $\hat{Y}$ is an approximately Gaussian variable we are comparing two Gaussian distributions with given means and variances. 

Having only access to estimators of noise and signal spectra ensemble averages, we use them to replace correlations in the test statistics.
Consequently, the average of $\hat{Y}$ (i.e. the optimal statistics using estimators for noise correlations) under ${\cal H}_{0}$ is given 
\begin{align}
	\mu_{{\cal H}_{0}} & = \frac{1}{2}\left[\check{\mathbb{C}}_{n}^{-1}\check{\mathbb{C}}_{g}\check{\mathbb{C}}_{n}^{-1}\right]_{ij}^{{\cal AB}} \langle{n_{i}^{{\cal A}}n_{j}^{{\cal B}}}\rangle\\
	& =\frac{1}{2}\text{Tr}\left(\check{\mathbb{C}}_{n}^{-1}\check{\mathbb{C}}_{g}\check{\mathbb{C}}_{n}^{-1}\mathbb{C}_{n}\right),\label{eq:muH0_G}
\end{align}
where the trace is performed over detector and data indices.
In a similar way we find, under the hypothesis ${\cal H}_{1}$
\begin{align}
	\mu_{{\cal H}_{1}} & =\mu_{{\cal H}_{0}}+\frac{1}{2}\text{Tr}\left(\check{\mathbb{C}}_{n}^{-1}\check{\mathbb{C}}_{g}\check{\mathbb{C}}_{n}^{-1}\mathbb{C}_{g}\right),
\end{align}
and the variances are given by
\begin{align}
	\sigma_{{\cal H}_{0}}^{2} & =\frac{1}{2}\text{Tr}\left[(\check{\mathbb{C}}_{n}^{-1} \check{\mathbb{C}}_{g} \check{\mathbb{C}}_{n}^{-1} \mathbb{C}_{n})^2\right] , \\
	\sigma_{{\cal H}_{1}}^{2} & \simeq\sigma_{{\cal H}_{0}}^{2}+\text{Tr}\left(\check{\mathbb{C}}_{n}^{-1}\check{\mathbb{C}}_{g}\check{\mathbb{C}}_{n}^{-1}\mathbb{C}_{n}\check{\mathbb{C}}_{n}^{-1}\check{\mathbb{C}}_{g}\check{\mathbb{C}}_{n}^{-1}
	\mathbb{C}_{g}\right) ,
\end{align}
where once again we included only the first correction for $\sigma_{{\cal H}_{1}}^{2}$ in the small signal approximation. 
The Receiver Operating Characteristic (ROC) of the DS reads
\begin{eqnarray}
	\label{eq:ROCgauss}
	P_\text{D}=\frac{1}{2} \text{erfc}\left(\frac{\sigma_{{\cal H}_{0}}}{\sigma_{{\cal H}_{1}}}\text{erfc}^{-1}(2P_\text{FA})-\frac{\mu_{{\cal H}_{1}}-\mu_{{\cal H}_{0}}}{\sigma_{{\cal H}_{1}}\sqrt{2}}\right)\label{eq:prob_detection},
\end{eqnarray}
where $P_\text{D}$ ($P_\text{FA}$) is the detection (false-alarm) probability, and $\text{erfc}$ is the complementary error function.

Note that, because of the two traces over the $N$ data points $(\mu_{{\cal H}_{1}}-\mu_{{\cal H}_{0}})/\sigma_{{\cal H}_{1}}\propto\sqrt{N}$, so the detection probability improves with the square root of the measurement time.
On the contrary, the first term affects only mildly the detector performance as it remains constant while more data points are accumulated. 
For this reason, Eq.~\eqref{eq:prob_detection} is often rewritten with trivial definitions for the ``offset'' \textit{o} and ``deflection coefficient'' \textit{d} as follows:
\begin{eqnarray}
	P_\text{D}=\frac{1}{2} \text{erfc}\left(\textit{o}-\sqrt{d^2}\right) .\label{eq:prob_detection_short}
\end{eqnarray}

However, the approach just illustrated is not always viable: to attain a detection we need to know $\mu_{{\cal H}_{0}}$ with an error of the order of the ratio between the signal's and the noise's power spectra. 
This is because we need to know $\mu_{{\cal H}_{1}}-\mu_{{\cal H}_{0}}$ with the same precision.
This cannot be done experimentally (we cannot switch off the coupling of the detectors to the SGWB) and it is not realistic to estimate theoretically the noise budget of a detector with such precision.

Usually this issue is solved by the additional assumption that noises across different detectors are uncorrelated, namely the matrix $\mathbb{C}_n^{\cal A B}$ is diagonal in detector's indices.
Consequently, the noise dominated terms along the diagonal ${\cal A}={\cal B}$ can be eliminated by defining a ``diagonal-free'' statistic $\hat{Y}_G(s)$ by removing in the sum of Eq.~\eqref{eq:Ygauss} all terms dominated by the noise
\begin{align}
	\label{eq:Ygaussdiagonalfree}
	\hat{Y}_G(s)
	\equiv 
	\sum_{\mathcal{A}\neq\mathcal{B}}
	\frac{1}{2} 
	\left[
	\check{\mathbb{C}}_{n}^{-1}
	\check{\mathbb{C}}_{g}
	\check{\mathbb{C}}_{n}^{-1}\right]_{ij}^{{\cal AB}} s_i^{\cal A}s_j^{\cal B} .
\end{align}
We get a new average $\mu_{{\cal H}_{0,\text{G}}}=0$ and the detector becomes robust with respect to errors in the noise model. The new means and variances are given by
\begin{align}
	\mu_{{\cal H}_{0,\text{G}}} & = 0 \, ,\label{eq:m0gauss} 
	\\
	\mu_{{\cal H}_{1,\text{G}}} & =
	\frac{1}{2}
	\text{Tr}
	\left(
	\check{\mathbb{C}}_{n}^{-1}
	{\check{\slashed{\mathbb{C}}}}_{g}
	\check{\mathbb{C}}_{n}^{-1}
	\mathbb{C}_{g}\right) ,
	\label{eq:m1gauss}
	\\
	\sigma_{{\cal H}_{0,\text{G}}}^{2} & =
	\frac{1}{2} \text{Tr}\left(
	(\check{\mathbb{C}}_{n}^{-1} \check{\slashed{\mathbb{C}}}_{g} \check{\mathbb{C}}_{n}^{-1} \mathbb{C}_{n})^2
	\right) ,
	\label{eq:sigma0gauss}\\
	\sigma_{{\cal H}_{1,\text{G}}}^{2} & \simeq \sigma_{{\cal H}_{0,\text{G}}}^{2}\! +\text{Tr}\left(
	\check{\mathbb{C}}_{n}^{-1}\check{\slashed{\mathbb{C}}}_{g}\check{\mathbb{C}}_{n}^{-1}\mathbb{C}_{n}\check{\mathbb{C}}_{n}^{-1}
	\check{\slashed{\mathbb{C}}}_{g}
	\check{\mathbb{C}}_{n}^{-1}
	\mathbb{C}_{g}
	\right) ,
	\label{eq:sigma1gauss}
\end{align}
where we label diagonal\textendash free correlation matrices (with no implicit summation over detector indices)
\begin{equation}
	\left[\slashed{\mathbb{C}}\right]^{{\cal A}{\cal B}} = \left[{\mathbb{C}}\right]^{{\cal A}{\cal B}}\left(1-\delta^{{\cal A}{\cal B}}\right) .
\end{equation}

Notably $\mu_{{\cal H}_{1,\text{G}}}-\mu_{{\cal H}_{0,\text{G}}}<\mu_{{\cal H}_{1}}-\mu_{{\cal H}_{0}}$ based on Eq.~\eqref{eq:prob_detection}: the additional robustness introduced affects the deflection coefficient, i.e. its asymptotic performances.

\subsubsection{Non-Gaussian case\label{subsubsec:non-Gaussian}}

In the more general case of a non-Gaussian model, we rewrite Eq.~\eqref{eq:Ystat} as
\begin{equation}\label{eq:statnongauss}
	\hat{Y}(s)\!=\!\frac{1}{2} [\check{\mathbb{C}}_g]^{\cal A B}_{i j} \mathfrak{s}_{i}^{\cal A} \mathfrak{s}_{j}^{\cal B}
	\!+\!\chi_h\! \sum_{n=1}^{\infty}\frac{1}{n!}\check{\Gamma}_{i_{1}\cdots i_{n}}^{{\cal A}_{1}\cdots{\cal A}_{n}} \mathfrak{s}_{i_{1}}^{{\cal A}_{1}}\cdots\mathfrak{s}_{i_{n}}^{{\cal A}_{n}}
\end{equation}
with
\begin{equation}
	\chi_h = \left\langle e^{-\frac{1}{2}{\cal W}_{n+g}\left(h,h\right)}\right\rangle .
\end{equation}
Here $\mathfrak{s}_{i}^{{\cal A}}=[\check{\mathbb{C}}_n^{-1}]_{ij}^{{\cal AB}}s_{j}^{\cal B}$ is a ``double whitened" signal, and $\check{\Gamma}_{i_{1}\cdots i_{n}}^{{\cal A}_{1}\cdots{\cal A}_{n}}$ are estimators of the connected moments for an $h$ distributed according to
\begin{align}
	p_{h}^{\prime}[h]=\chi_h^{-1} e^{-\frac{1}{2}{\cal W}_{n+g}\left(h,h\right)}p_{h}[h] 
\end{align}
and are fully contracted over a suitable number of signals $\mathfrak{s}_{i}^{\mathcal{A}}$, which we denote with a subscript $\left\lbrace \mathcal{A},i\right\rbrace$.

Now we can evaluate the expectation value of $\hat{Y}(s)$ under the hypothesis ${\cal H}_{0}$. 
We find
\begin{align}
	\label{eq:mu0nongauss}
	\mu_{{\cal H}_{0}}
	=&\frac{1}{2}\text{Tr}\left(\check{\mathbb{C}}_{n}^{-1}\check{\mathbb{C}}_{g}\check{\mathbb{C}}_{n}^{-1}\mathbb{C}_{n}\right) \nonumber\\
	+&\chi_h \sum_{n=1}^{\infty}\frac{1}{n!}\check{\Gamma}_{i_{1}\cdots i_{n}}^{{\cal A}_{1}\cdots{\cal A}_{n}} 
	\mathbb{N}_{i_{1}\cdots i_{n}}^{{\cal A}_{1}\cdots {\cal A}_{n}} .
\end{align}
where $\mathbb{N}_{i_{1}\cdots i_{n}}^{{\cal A}_{1}\cdots {\cal A}_{n}}$ is the $n$--th order moment of the double whitened noise.
As the noise is Gaussian, by virtue of Isserlis theorem~\cite{10.2307/2331932} it can be written as a sum over all pairings of products of second order moments, and using the simmetry of the connected moments $\Gamma$ we get

\begin{align}
	\label{eq:mu0nongaussB}
	\mu_{{\cal H}_{0}}=& \frac{1}{2}  \text{Tr}\left(\check{\mathbb{C}}_{n}^{-1}\check{\mathbb{C}}_{g}\check{\mathbb{C}}_{n}^{-1}\mathbb{C}_{n}\right) \nonumber \\ 
	+& \chi_h\!\! \sum_{n=1}^{\infty}\!\frac{1}{(2n)!!}\check{\Gamma}_{i_{1}\cdots i_{2n}}^{{\cal A}_{1}\cdots{\cal A}_{2n}} 
	\prod_{k=1}^n 
	[\check{\mathbb{C}}_n^{-1} \mathbb{C}_n \check{\mathbb{C}}_n^{-1}]_{i_{(2k-1)} i_{2k}}^{ {\cal A}_{(2k-1)} {\cal A}_{2k}} .
\end{align}

As for the Gaussian case, $\mu_{{\cal H}_{0}}$ depends on an estimate of the real spectral covariance of the noise, which is not sufficiently under control. 
We can set to zero the first term in Eq.~\eqref{eq:mu0nongaussB} by using the same approach discussed for the Gaussian case, but this is not enough to eliminate the second. 
In order to obtain a robust detector we define the new statistic
\begin{equation}
	\label{eq:scrambled}
	\mathring{Y}(s) \equiv \hat{Y}(s)-\hat{Y}(\mathring{s}).
\end{equation}
Here $\mathring{s}$ are the observed data, transformed in such a way that $\mathring{s}^{\cal A}_i$ satisfies the following:
\begin{align}
	\left\langle
	\mathring{s}^{\cal A}_i
	\mathring{s}^{\cal B}_j\right\rangle 
	& = \delta^{\cal A B} 
	\left[ \mathbb{C}_{s} \right]_{ij}^{{\cal AB}} , \label{eq:AA} \\
	\left\langle
	\mathring{s}^{\cal A}_i
	s^{\cal A}_j
	\right\rangle \label{eq:BB}
	& = 0 \, .
\end{align}

This can be done by introducing appropriate and large enough shifts among detectors' data in time-domain \cite{2010CQGra..27a5005W}, randomizing the phases in frequency domain, or scrambling data chunks, such that the original series of each detector and the new ones are statistically independent [therefore implying Eq.~\eqref{eq:BB}], and the cross-correlations across detectors are removed [i.e. Eq.~\eqref{eq:AA}]. Henceforth, we will denote $\mathring{s}$ and $\mathring{Y}(s)$ -- the latter not to be confused with $\hat{Y}(\mathring{s})$,  a statistics insensitive by construction to the GW signal -- as ``scrambled data'' and ``scrambled detection statistic'', respectively.
We defer a detailed characterization of the statistical subtleties of this procedure in a realistic scenario to future study. 

Under hypothesis ${\cal H}_0$ the correlations are computed on noise--only data, therefore they are already diagonal in detector's indices, so $\langle{\hat{Y}(s)}\rangle=\langle{\hat{Y}(\mathring{s})}\rangle$ and we get 
\begin{align}\label{eq:m0nongaussianshift}
	\mathring{\mu}_{{\cal H}_0}\equiv \langle{\mathring{Y}(s)}\rangle = 0. 
\end{align}

Taking the expectation value under the hypothesis ${\cal H}_1$ we find

\begin{align}\label{eq:m1nongaussianshift}
	\mathring{\mu}_{{\cal H}_{1}} & = 
	\frac{1}{2} \text{Tr}\left(
	\check{\mathbb{C}}_{n}^{-1}
	\check{\slashed{\mathbb{C}}}_{g}
	\check{\mathbb{C}}_{n}^{-1}
	\mathbb{C}_{g}
	\right) 
	+ \frac{1}{2} \text{Tr}\left(
	\check{\mathbb{C}}_{n}^{-1}
	\check{\slashed{\mathbb{C}}}_{g}
	\check{\mathbb{C}}_{n}^{-1}
	\mathbb{C}_{h}
	\right) \nonumber\\
	&+  \chi_h \sum_{n=1}^{\infty}\frac{1}{n!}\check{\Gamma}_{i_{1}\cdots i_{n}}^{{\cal A}_{1}\cdots{\cal A}_{n}} 
	\left( \mathbb{S}_{i_{1}\cdots i_{n}}^{{\cal A}_{1}\cdots {\cal A}_{n}} - \mathring{\mathbb{S}}_{i_{1}\cdots i_{n}}^{{\cal A}_{1}\cdots {\cal A}_{n}}  \right),
\end{align}
where $\mathbb{S}$ are the momenta of the signal $\mathfrak{s}$ and $\mathring{\mathbb{S}}$ the momenta of $\mathring{\mathfrak{s}}$.  

When $h=0$  only the first term is non-null, reproducing the Gaussian result (on scrambled data). From the sum we see that additional contributions arise in the general case.
These are expected to improve the DS performances, and open up the possibility of a stricter characterization of the SGWB statistical properties. 

\subsection{\label{subsec:Bayesian-approach}The Bayesian approach}

The detection and parameter estimation proposed in Sec.~\ref{subsec:NPdetector} can be equivalently formulated in a Bayesian context observing that Eq.~\eqref{eq:SignalProbability} is the unnormalized probability distribution of the observed data conditioned on a given model, i.e. the likelihood.
The probability distribution of a model given some observed data (i.e. the posterior) is obtained through Bayes theorem, as
\begin{equation}
	\label{eq:posterior}
	p\left({\cal M}|s\right)\!\propto\! {\cal N}_{n+g}\!\!\int_h \!\!e^{-\frac{1}{2}{\cal W}_{n+g}(s-h,s-h)} p_h[h|{\cal M}] \pi({\cal M}) ,
\end{equation}
where ${\cal M}$ and $\pi({\cal M})$ are the model and its prior distribution, up to a model independent normalization constant.

The posterior can be estimated with importance sampling using a Monte Carlo Markov-chain (MCMC), by generating a sequence of ${\cal M}$'s with the probability distribution defined by Eq.~\eqref{eq:posterior}.
In principle each MCMC step would require a non trivial integration to be performed, and this can also be obtained with a nested importance sampling.
This is a nontrivial task, as the estimation of the integral in Eq.~\eqref{eq:posterior} has statistical errors roughly proportional to $1/\sqrt{N_s}$, where $N_s$ is the number of the evaluation steps. 
A trade-off between the amount of knowledge on the probability distribution and the computational cost of the procedure would be required.

A better procedure can be devised by focusing on the integrand of Eq.~\eqref{eq:posterior}, i.e. the joint posterior on model and non-Gaussian realization $h$:
\begin{equation}
	\label{eq:posteriorfull}
	\!\!\!p\left(h,{\cal M}|s\right)\!\propto {\cal N}_{n+g} e^{-\frac{1}{2}{\cal W}_{n+g}(s-h,s-h)} p_h[h|{\cal M}] \pi({\cal M}) ,
\end{equation}
a single sequence of ${\cal M}$'s and $h$'s can be generated at the same time with MCMC techniques. 
This does not solve the computational cost issue, but makes evident that in principle the model estimation can be improved at will with a large enough number of MCMC steps.
Each step can be performed along the lines of the Metropolis--Hastings algorithm as follows:
\begin{description}
	\item[Step $1$] Starting from a model ${\cal M}_k$, a new one ${\cal M}_{k+1}$ is generated with transition distribution $T({\cal M}_{k+1}\mid {\cal M}_{k})$.
	\item[Step $2$\namedlabel{it:2}{2}] A realization $h_{k+1}$ is generated accordingly with the distribution $p_h[h_{k+1}|{\cal M}_{k+1}]$. In Sec.~\ref{subsec:Importance-sampling} we provide a well defined procedure for this purpose.
	\item[Step $3$\namedlabel{it:3}{3}] The value of 
	\begin{equation}
		\mathcal{I}_{k+1}={\cal N}_{n+g} e^{-\frac{1}{2}{\cal W}_{n+g}(s-h_{k+1},s-h_{k+1})}
	\end{equation} is compared with the one evaluated at the previous step, and the new model is accepted with probability 
	\begin{equation} \label{eq:acceptance}
		\min \left\{ 1, \frac{\mathcal{I}_{k+1}}{\mathcal{I}_k} \frac{T({\cal M}_{k+1},{\cal M}_{k})}{T({\cal M}_{k},{\cal M}_{k+1})}\right\}.
	\end{equation}
	Otherwise the process is repeated.
\end{description}

Using this approach the prior probability $\pi({\cal M})$ is not considered, and can be used later to obtain the posterior.
Depending on the framework chosen, the model $\cal M$ can be constructed to explore a fixed-dimension parametric family of distributions, or can be tailored to explore models with different dimensions using reversible jump MCMC methods~\cite{10.1093/biomet/82.4.711}.
In addition, $\pi({\cal M})$ can be incorporated simply with the redefinition
\begin{equation}
	\mathcal{I}_{k+1}={\cal N}_{n+g} e^{-\frac{1}{2}{\cal W}_{n+g}(s-h_{k+1},s-h_{k+1})} \pi({\cal M}_{k+1})
\end{equation}
This can be an advantage in some specific cases when the prior is informative and optimization of convergence rate is required.

In a generalized approach, Steps~\ref{it:2} and~\ref{it:3} are modified as follows:
\begin{description}
	\item[Step $2^\prime$] A set of $N_s$ sequences $h_{k+1,i}$ are generated accordingly with the distribution $p_h[h_{k+1,i}|{\cal M}_{k+1}]$.
	\item[Step $3^\prime$] The value of
	\begin{equation}
		\mathcal{I}_{k+1}\equiv\frac{1}{N_s} \sum_i {\cal N}_{n+g} e^{-\frac{1}{2}{\cal W}_{n+g}(s-h_{k+1,i},s-h_{k+1,i})}
	\end{equation}
	is compared with the one evaluated at the previous step, and the new model is accepted or rejected with the same rule described in Eq.~\eqref{eq:acceptance}.
\end{description}

The likelihood in Eq.~\eqref{eq:posterior} is obtained in the limit $N_s\rightarrow~\!\!\!\infty$, and we can see $N_s$ as a free parameter to tune. 

The MCMC sampler favours models with a low value of ${\cal W}_{n+g}(s-h,s-h)$, and this can be interpreted as follows: new models are accepted at each MCMC steps when they perform better at removing the non-Gaussian part of the signal. When such part is weak we can expect that a decision based on a single sequence $h$ could be dominated by statistical fluctuations, so averaging over a large value of $N_s$ provides additional robustness to the algorithm. 

Note that the convergence of the sampler is guaranteed for each value of $N_s$.
\begin{figure}
	\centering
	\includegraphics[width=0.95\columnwidth]{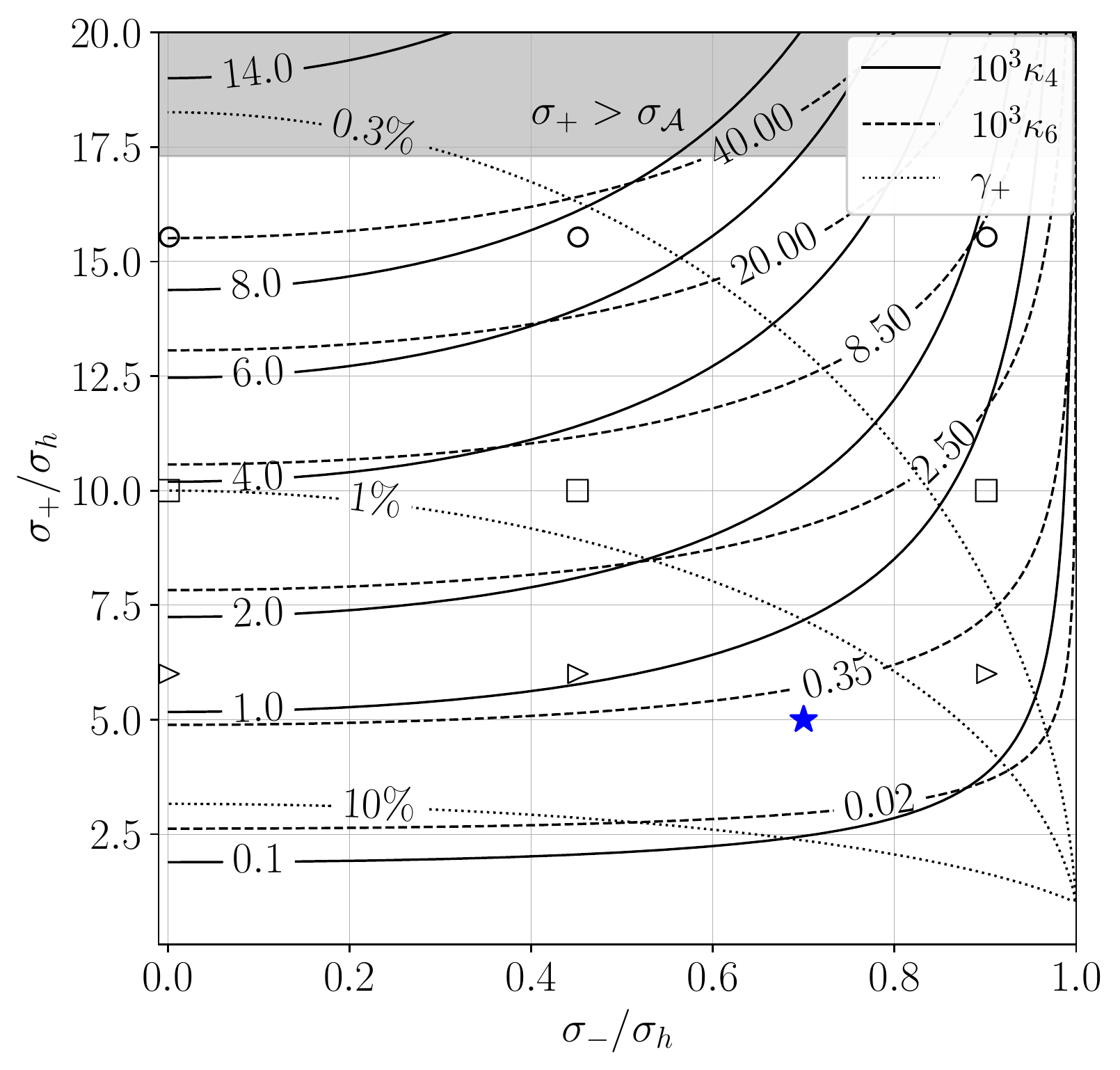}
	\caption{Cumulants values for the mixture toy model introduced in Sec.~\ref{sec:toymodel} over the full parameter space. Solid and dashed lines denote mixtures with equal fourth-- and sixth--order cumulants, $k_4$ and $k_6$ respectively. 
		Dotted black lines denote models with equal mixture weights $\gamma_+$. The shaded gray region denotes models with the brightest of the two components, $\gamma_+$ greater than the noise level of a single detector $\sigma_{\mathcal{A}}$. Any given two black lines intersect only once, hence providing an alternative representation of the full parameter space. 
		Circles (squares, triangles) denote a discrete set of models with various levels of non-Gaussianity. Their DSs is characterized in greater details, with results and signal realizations shown in top (middle, bottom) panels of Fig~\ref{fig:ROCtoyGrid}.
		The blue star denotes an additional model, exhibiting significant correlations between $\sigma_+$ and $\sigma_-$. We use this model to characterize the performance of a Bayesian parameter estimation, as described in Sec.~\ref{sub:toybayes}. 
		As shown by the posterior in Fig.~\ref{fig:bayesianposterior}, parametrizing the mixture model through its cumulants helps naturally decorrelate them.}
	\label{fig:cumul}
\end{figure}

As we shall see in Sec.~\ref{sec:toymodel} (see Fig.~\ref{fig:cumul},~\ref{fig:ROCtoy} and~\ref{fig:bayesianposterior}), the natural parameterization of the non-Gaussian  components in terms of higher-order cumulants, along the lines of the Edgeworth or Gram--Charlier A expansions~\cite{PhysRevD.89.124009, Racine:2007gv},  appears to also fit in a Bayesian context as it provides parameters inherently decorrelated upon inference.
In Sec.~\ref{subsec:point-process} we show how this is also a very convenient parametrization for SGWBs characterized by the incoherent superposition of multiple independent signals, with a significant reduction of the computational cost to perform importance sampling.
This is a crucial need of our proposed algorithm: Eq.~\eqref{eq:SignalProbability} is a sort of ``Wiener filter'' with stochastic templates. If their sample space is complicated to explore, e.g. when the duty cycle~\cite{Regimbau:2011rp} of the background is low, a large number of evaluations might be needed to ensure the algorithm convergence.

\section{A toy model example}
\label{sec:toymodel}

Let us consider the very simplified model
\begin{equation}
	s^{{\cal A}}_i = h^{{\cal A}}_i+n^{{\cal A}}_i ,
\end{equation}
The noise is modeled by uncorrelated Gaussian variables $n^{\cal A}_i$  with 
\begin{align}
	\langle{n_i^{\cal A}}\rangle&=0 ,\\
	\langle{n_i^{\cal A} n_j^{\cal B}}\rangle&=\sigma_{\cal A}^2 \delta^{\cal A B} \delta_{ij} \, .
\end{align}
We set $h_i^\mathcal{A}=h_i$,  where $h_i$ are independent variables with probability distribution 
\begin{align}
	\label{eq:prob-of-h}
	p(h_i) & =\gamma_+ \mathcal{N}(h_i;\sigma_+) + \gamma_- \mathcal{N}(h_i;\sigma_-) ,  \\
	\gamma_+ & = \frac{\sigma_h^2-\sigma_-^2}{\sigma_+^2-\sigma_-^2}\,, \qquad
	\gamma_-  = 1-\gamma_+ = \frac{\sigma_+^2-\sigma_h^2}{\sigma_+^2-\sigma_-^2} .
\end{align}

Here $\mathcal{N}(x;\sigma_i)$ is a Gaussian distributions for $x$ with zero mean and variance $\sigma_i^2$, and the parametrization of ordered variances $\sigma_+^2>\sigma_h^2>\sigma_-^2$ is chosen such that for any values of $\sigma_+$, $\sigma_-$ the distribution variance is $\sigma_h^2$. The kurtosis is given by 
\begin{equation}\label{eq:kurtosis}
	\beta\equiv 
	\left\langle 
	\frac{h_i^4}{\sigma_{h}^4}
	\right\rangle
	=
	3
	\left(
	\frac{\sigma_{+}^{2}}{\sigma_{h}^{2}} 
	-\frac{\sigma_{+}^{2}}{\sigma_{h}^{2}}
	\frac{\sigma_{-}^{2}}{\sigma_{h}^{2}} 
	+\frac{\sigma_{-}^{2}}{\sigma_{h}^{2}} 
	\right) ,
\end{equation}
which has a minimum of $3$ when $\sigma_+=\sigma_h$ or $\sigma_-=\sigma_h$ (Gaussian cases with $\gamma_{+,-}=1$) and grows larger and larger with $\sigma_+$. 
The whole family of leptokurtic probabilities, parameterized by $\sigma_+,\sigma_-,\sigma_h$ can be equivalently explored by three nontrivial cumulants $k_n$, formally defined by the power expansion of the cumulant generating function $K$ (see Appendix~\ref{sec:cumulantstoymodel} for more details)
\begin{align}
	K(t)& =
	\log 
	\left\langle 
	e^{t X}
	\right\rangle \\
	k_n & = 
	\left.
	\frac
	{\partial^n K(t)}
	{\partial t^n}
	\right|_{t=0},\, 
	n=2,4,6 \, .
\end{align}
For our toy model they are equal to
\begin{align}
	k_2\!&= \sigma_h^2 ,\\
	k_4\! &= 
	3 (\sigma_-^2\sigma_h^2 
	+ \sigma_+^2\sigma_h^2
	- \sigma_+^2\sigma_-^2
	-\sigma_h^4 
	) ,
	\\
	k_6\! &=
	15 \left(\sigma_+^2-\sigma_h^2\right) \left(\sigma_-^2-\sigma_h^2\right) 
	\left(2\sigma_h^2-\sigma_+^2-\sigma_-^2\right) .
\end{align}

In Fig.~\ref{fig:cumul} we plot contours of constant cumulants $k_4,k_6$ as a function of the mixture parameters $\sigma_+,\sigma_-$, at a reference value of $\sigma_h$, alongside the mixture component weights, uniquely specified by $\gamma_+$.  
It is apparent that the non-linear relation between $k_i$ and $\sigma_{\pm}$ could affect significantly the stochastic sampling involved in the Bayesian analysis, while for a frequentist DS it serves only as an alternative parametrization.

Though very simple, this model is expected to capture some features of a realistic non-Gaussian background. For example, the particular case $\sigma_-=0$ represents backgrounds with burst--like events which are so short that their structure cannot be resolved. 
One of them (and only one) can be present or not at a given time with a specific probability $\gamma_+$, and their amplitude has a Gaussian distribution with standard deviation $\sigma_+$, somewhat in the spirit of the analysis in~\cite{Smith:2018PhRvX...8b1019S}.
As only a single event can contribute to the signal at a given time, statistical independence holds: $P(h(t_1),\cdots h(t_k))= \prod_k P(h(t_k))$. 
In a realistic scenario this is not true. Assuming the event waveform has a given shape $u_i$, the strain at a given time contains contributions from several events. 
In some peculiar cases it is possible to factorize the probability distribution by using a different domain to describe the signal (e.g. frequency for monochromatic waveforms) but this will be impossible in a generic setup, and the full machinery of point processes \cite{CoxIsham1980} described in Sec.~\ref{subsec:point-process} should instead be adopted.

\subsection{\label{sub:toyNP}Frequentist detection}

The DS in Eq.~\eqref{eq:Ystat} can be evaluated analytically for the chosen toy model. 
As the noise spectrum is white and the signal values across different data points are independent we have (see  Eq.~\eqref{eq:tmYstat} for a detailed proof)
\begin{align}\label{eq:toyNPstatistic}
	\hat{Y}(s) = \sum_i \log \left< \exp \left[ -\sum_{\cal A}
	\frac{h_i(h_i-2s_i^{\cal A})}{2\sigma_{\cal A}^2}  \right] \right> .
\end{align}

The expectation value can be evaluated explicitly, obtaining
\begin{equation}
	\hat{Y}(s)  = \sum_{i} \hat{y}\left[u\left(s_{i}\right)\right]\label{eq:toyNPstatisticB}
\end{equation}
with $\hat{y}$ a non trivial function of a single datapoint
\begin{align}
	\label{eq:optimalY}
	\hat{y}(u)  
	& = \!
	\log\!\left[ 
	\sum_{\alpha=+,-}
	\!\frac{\gamma_\alpha \sigma }{\sqrt{\sigma^2+\sigma_\alpha^2}}
	\exp\!
	\left(
	\frac{\sigma_\alpha^2 u^2}{2(\sigma^2+\sigma_\alpha^2)}
	\right)
	\!
	\right] ,\\
	u\left(s_{i}\right)
	& = 
	\sigma 
	\sum_{\mathcal{A}} 
	\frac{s_{i}^{\mathcal{A}}}{\sigma_{\mathcal{A}}^{2}} ,\\
	\frac{1}{\sigma^2} 
	& \equiv\sum_{\cal A} \frac{1}{\sigma_{\cal A}^2} .\label{eq:effective-noise}
\end{align}

When the number of datapoints is large $\hat{Y}$ becomes a Gaussian variable according to the central limit theorem, so mean and variance suffice to characterize the detection performances. 

Under the hypothesis ${\cal H}_0$ the variable $u$ is by definition normally distributed. 
\begin{equation}
	p(u) \distas{{\cal H}_0} \mathcal{N}\left(u;1\right) .
\end{equation}

Under the hypothesis ${\cal H}_1$ the expectation value of $u$ is still zero, but the variance gets an additive contribution from the signal. For unscrambled data, we get
\begin{equation}
	p(u) \distas{{\cal H}_1} \gamma_+ \mathcal{N}\!\left(u;\sqrt{1+ \frac{\sigma_+^2}{\sigma^2}}\right) + \gamma_- \mathcal{N}\left(u; \sqrt{1+ \frac{\sigma_-^2}{\sigma^2}}\right)
\end{equation}
while the equivalent formula for scrambled data is discussed in Appendix~\ref{app:toy-model-derivation}.

\begin{figure*}
	\centering
	\includegraphics[width=\textwidth]{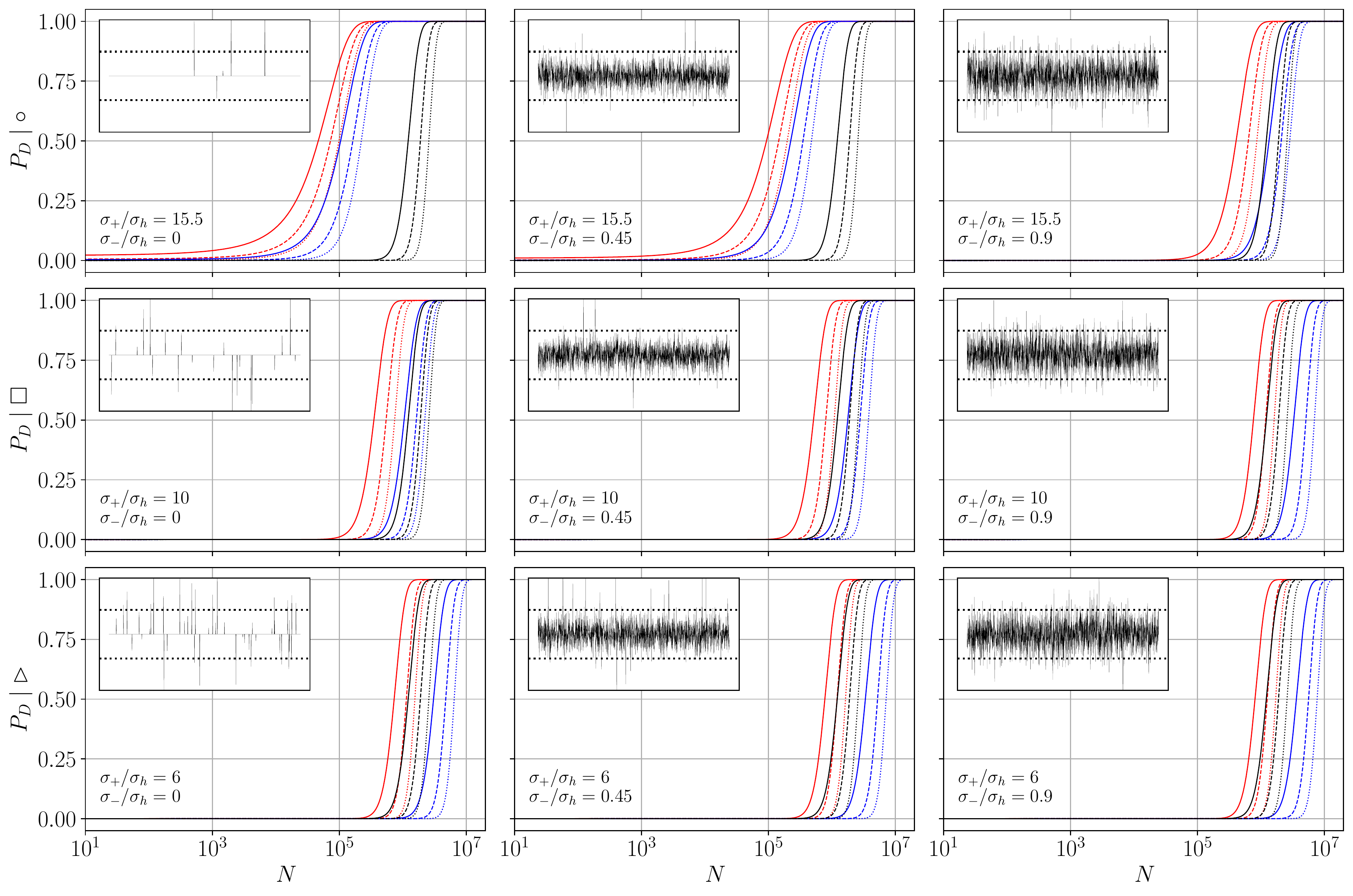}
	\caption{
		Performances comparison between DSs for a selection of models across the parameter space in Fig.~\ref{fig:cumul}. 
		Black (red, blue) lines denote the probability of detection $P_D$ as a function of the number of datapoints $N$ for the Gaussian (``optimal``, non-Gaussian on scrambled data) DS, i.e. $Y_G$ ($Y$, $\mathring{Y}$).	
		Solid (dashed, dotted) lines corresponds to a probability of false alarm $P_{FA}=10^{-10}$ ($10^{-15}$, $10^{-20}$). 
		The level $\sigma_h=0.1$ is kept constant for all models, resulting in an overall shift of the black curves. 
		Values for $\sigma_+/\sigma_h$ and $\sigma_-/\sigma_h$ are specified in each plot.
		Top (middle, bottom) row corresponds to models identified with circles (squares, triangles) in Fig.~\ref{fig:cumul} where higher order cumulants values can be recovered.
		Performances improve as the non-Gaussianity is enhanced, (top-left panel).
		The non-Gaussian DS on unscrambled data (red lines) outperforms the Gaussian one everywhere in the parameter space, and it performs similarly to it only for signals with small non-Gaussianity (bottom row, corresponding to triangles in Fig.~\ref{fig:cumul}).
		Data scrambling (blue lines) can suppress the advantage of the optimal non-Gaussian DS (red lines) if non-Gaussianity is not high enough.
		Upper left insets in each subplot show a short signal realizations for the respective model in absence of noise. For reference, detector noise levels $\pm\sigma_{\mathcal{A}}$ are shown as horizontal dashed black lines.
	}
	\label{fig:ROCtoyGrid}
\end{figure*}

In both cases we rewrite Eq.~\eqref{eq:ROCgauss},
\begin{eqnarray}
	\label{eq:ROCgauss2}
	P_\text{D}=\frac{1}{2} \text{erfc}\left(r_1 \text{erfc}^{-1}(2P_\text{FA})-d_1 \sqrt{\frac{N}{2}} \right),
\end{eqnarray}
where $r_1=\sigma_{{\cal H}_{0}}/\sigma_{{\cal H}_{1}}$ and $d_1=(\mu_{{\cal H}_{1}}-\mu_{{\cal H}_{0}})/\sigma_{{\cal H}_{1}}$ can be evaluated easily by numerical integration in the $N=1$ case.

In Fig.~\ref{fig:ROCtoyGrid} we show the performance of our DS for a discrete set of toy model parameters with various levels of non-Gaussianity. Circles, squares, and triangles identify sets of models with constant $\sigma_+$ and varying $\sigma_-$. We illustrate the detection probability $P_D$ as a function of the number $N$ of data points, alongside the respective signal realizations. We do this for three reference false alarm probabilities, and both for original and scrambled data. 

For comparison, we also show the performance of a Gaussian diagonal-free DS, namely
\begin{align}
	\hat{Y}_G(s) = \sum_i \sum_{{\cal A}\neq{\cal B}} \frac{s_i^{\cal A} s_i^{\cal B}}{\sigma_{\cal A}^2 \sigma_{\cal B}^2},
\end{align}
applied to the toy model data. The values of $r_1$ and $d_1$ for this particular case are evaluated in Appendix~\ref{app:toy-model-derivation}.

As expected, the non-Gaussian DS (without scrambled data) outperforms the scrambled and Gaussian ones.
However as discussed previously the optimal, non scrambled DS cannot be practically implemented.
The relevant performances to look at are those of the scrambled one. 
It performs better than the Gaussian one for large enough values of $k_4$ and $k_6$ (see Fig.~\ref{fig:cumul} where the set of parameters choosen for Fig.~\ref{fig:ROCtoyGrid} is shown). The Gaussian DS being better than the scrambled one for small non-Gaussianity is not unexpected: when we evaluate the scrambled statistics $\mathring{Y}$ we subtract two sets of data [see Eq.~\eqref{eq:scrambled}]
 in order to have zero average under ${\cal H}_0$. We pay a price for this, introducing additional fluctuations: the variance of the scrambled DS is the sum of the variances evaluated on normal and scrambled data. For small enough values of non-Gaussianity this price is larger than the gain obtained.  

Further insight is obtained by introducing a measure of the improvement between $\mathring{Y}$ and $\hat{Y}_G$.
 A simple possibility is to solve Eq.~\eqref{eq:ROCgauss2} for $N$, obtaining $N=N(P_{FA},P_D,r_1,d_1)$ for a given DS. We evaluate the ratio $N_G/\mathring{N}$ for fixed values of $P_D$ and $P_{FA}$ in the space of toy models' parameters. This is a measure of how much more data one needs to collect to achieve with the Gaussian DS performances similar to those of the scrambled one.
\begin{figure}
	\centering
	\includegraphics[width=\columnwidth]{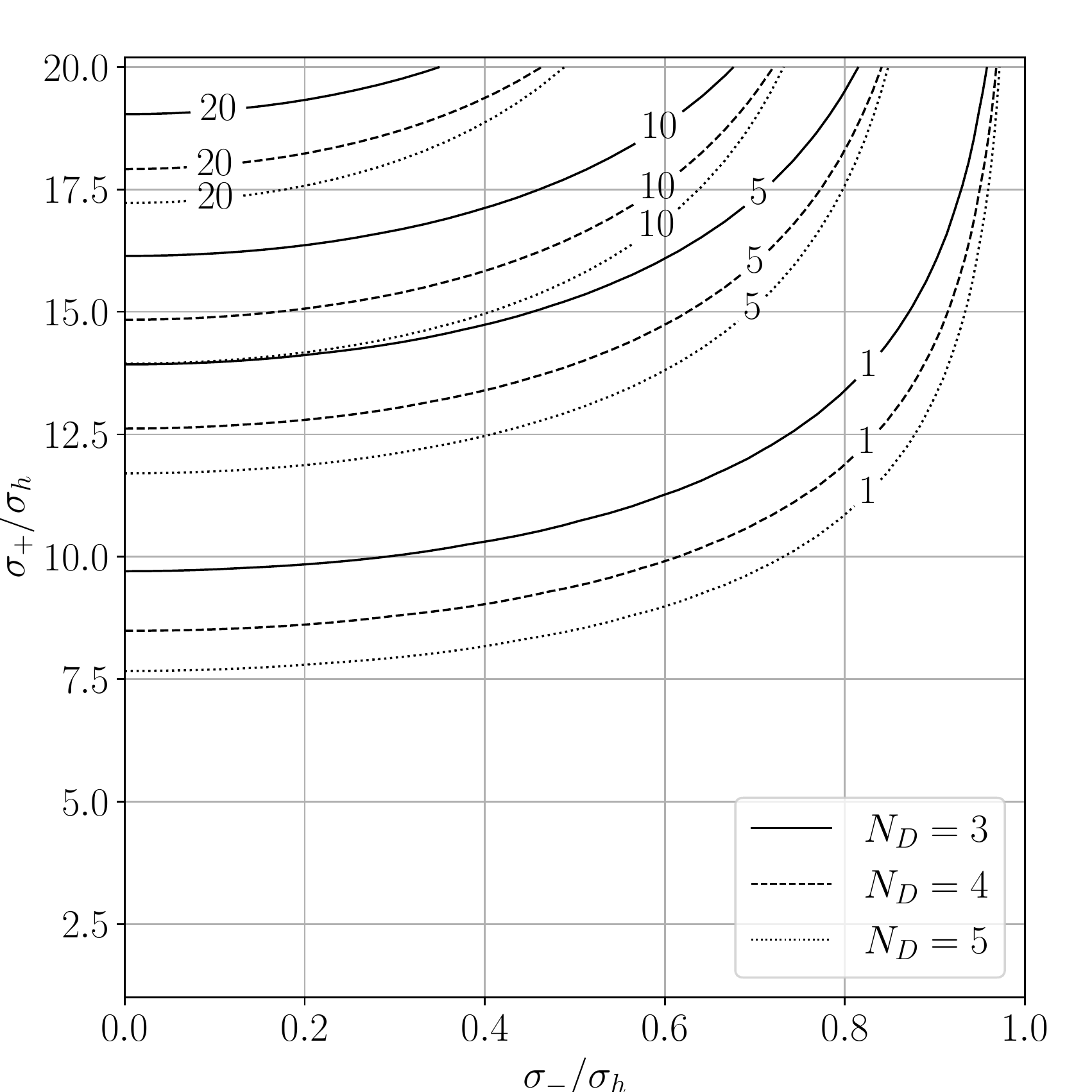}
	\caption[ROCshort]{
		Performance comparison between the Gaussian DS and the non-Gaussian one on scrambled data, shown across the toy model parameter space. We show contour levels of $N_G/\mathring{N}$, the number of datapoints required to achieve the same $P_D$ at a fixed $P_{FA}$. Black, red and blue solid lines denote contours for a configuration with three, four and five detectors, respectively.
		In the high non-Gaussianity limit (top-left corner) the Gaussian DS requires as many as twenty times more data to achieve comparable performances to the non-Gaussian one on scrambled data. 
		On the contrary, for mild non-Gaussianities the two DSs perform similarly. 
		In comparison, increasing the number of detectors improves the non-Gaussian DS performances moderately.
	}
	\label{fig:ROCtoy}
\end{figure}
The result is shown in Fig.~\ref{fig:ROCtoy}. It is evident that a significant advantage can be obtained in the large--non Gaussianity regime. 
We plot our results for different number of detectors in the network, and we observe that large $N_D$ gives improved performance of the scrambled statistics: this is expected because additional fluctuations introduced by the scrambled data do not scale with $N_D$.

It is clear that the scrambled data subtraction procedure is not optimal, and it is worth exploring alternative options. For example, the cumulant expansion in Eq.~\eqref{eq:statnongauss} could be used to define the generalization of a diagonal-free Gaussian DS, by removing terms not enough under control order-by-order, i.e. with non-zero expectation value under ${\cal H}_0$. This approach is useful especially in the mild non-Gaussian regime, where a truncation in the cumulant expansion is accurate enough. We leave this study to future investigation.

\begin{figure}
	\centering
	\includegraphics[width=\columnwidth]{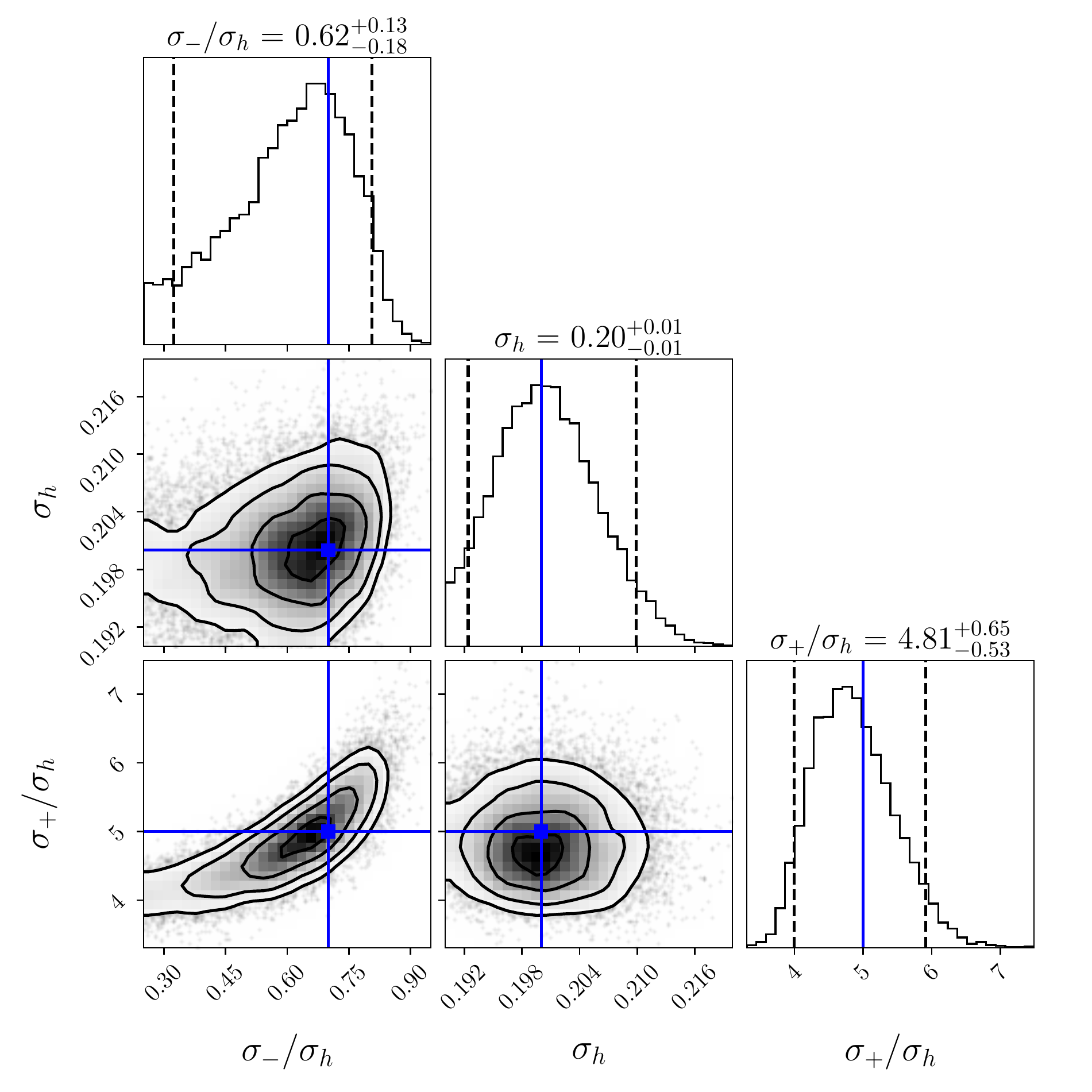}
	\caption[Bayes]{
		Corner plot of the Bayesian posterior for the toy model analysis. The noise level is set to $\sigma_{\cal A}=\sqrt{3}$ with $N_D=3$, is assumed known upon inference, and is additive to the signal in the data. 
		The number of samples is set to $N=4\times 10^5$ which grants the likelihood significant constraining power on the model parameters within the chosen prior.
		The true signal parameters are shown with solid blue lines, and correspond to the point in Fig.~\ref{fig:cumul} labelled with a star symbol.
		Black dashed lines denote posterior $90\%$ confidence intervals.
		Priors are uniform for all parameters, and relative ordering is enforced through hypertriangulation~\cite{PhysRevD.100.084041}. 
		The non-linear correlation observed in the bottom left subplot matches closely the levels of constant cumulants shown in Fig.~\ref{fig:cumul}, which suggest that the cumulant parametrization of the non-Gaussianities would be suitable for an efficient exploration of the parameter space.
		No predominance of a single cumulant can be identified in the posterior, as expected from contributions in~Eq.\eqref{eq:statnongauss}.}
	\label{fig:bayesianposterior}
\end{figure}

\subsection{\label{sub:toybayes}Bayesian parameter estimation}

The study of the Bayesian procedure with the toy model is simplified by the independence of noise and signal at different times.
Taking advantage of it we can write a recursive procedure which, given the posterior distribution for the model given $k$ data, evaluate the posterior distribution when we add the $k+1$ measurement.
As we show in Appendix~\ref{app:posterior_toymodel}, the likelihood (and subsequently the posterior) can be obtained analytically from Eq.~\eqref{eq:SignalProbability} and can be written as the product of likelihoods over individual data points. 
Explicitly, it reads
\begin{align}\label{eq:likelihood}
	\!\!\!\!\!\mathcal{L}\left(s_i \! \mid \! \mathcal{M}\right)\!\propto\!\!\!\!\!\sum_{\alpha=+,-}\!\! \frac{\gamma_\alpha }{\sqrt{1+\frac{\sigma^2_\alpha}{\sigma^2}}}e^{
		-\frac{1}{2} {\cal Q}_{\cal A B}^\alpha s_i^{\cal A} s_i^{\cal B}-\sum_{\cal A}\!\log \sqrt{2\pi} \sigma_{\cal A}
	},
\end{align}
where ${\cal Q}_{\cal A B}^\alpha$, proportional to the transverse projector in the detector space, is
\begin{align}\label{eq:QAB}
	{\cal Q}_{\cal A B}^\alpha & = \frac{1}{\sigma_{\cal A}^2 \sigma_{\cal B}^2} \left(\delta_{\cal A B}-\frac{\sigma_\alpha^2}{\sigma^2+\sigma_\alpha^2} \frac{\sigma_{\cal A}^{-1}\sigma_{\cal B}^{-1}}{\sigma^{-2}} \right).
\end{align}

In Fig.~\ref{fig:bayesianposterior} we show the results of an inference performed on a representative model (identified by a blue star in Fig.~\ref{fig:cumul}). We perform inference through stochastic nested sampling~\cite{10.1214/06-BA127}, using the software package \textsc{cpnest}~\cite{2021zndo...4470001V}. 
We use uniform priors for $\sigma_{-,+}$ and $\sigma_h$, and enforce their mutual ordering through hypertriangulation~\cite{PhysRevD.100.084041}.
The detector noise levels $\sigma_{\cal A}$ are fixed and assumed known.
For ease of comparison with Fig.~\ref{fig:cumul} we show posteriors and confidence intervals for the dimensionless parameters $\sigma_{+}/\sigma_h$ and $\sigma_{+}/\sigma_h$.

Notably, the two-dimensional posterior for $\sigma_{-,+}$ has most of its support along regions of constant cumulants.
This suggests that the cumulant parameterization of non-Gaussianities, beyond its naturalness in a statistical sense, is efficient at reducing correlations upon stochastic sampling of the parameter space.

\section{Application to astrophysical backgrounds}
\label{sec:AstroSB}

An important example of an SGWB exhibiting non-Gaussianity is that of astrophysical origin~\cite{2014PhRvD..89h4063M}. 
The stochastic signal can be modelled as the result of many uncorrelated events superpositions, each event contributing with a well-defined waveform (a function of the source parameters, predictable only in a statistical sense).
If there is a strong overlap between these contributions, in a sense that will be defined quantitatively below, the result is a Gaussian background. If this is not the case, non-Gaussian effects appear: the background is no more completely described by its power spectrum, and some additional modelling is required.

\subsection{\label{subsec:point-process}Point processes}

We parameterize the incoherent superposition of multiple signals as a stochastic process 
\begin{equation}\label{eq:camp}
	h_i^\mathcal{A}=\sum_{\sigma=1}^{N}u^\mathcal{A}_i(\theta_\sigma),
\end{equation}
where $N$ is a discrete random variable, describing the number of individual signals for a given realization of $h$.
$u_i^\mathcal{A}$ are effective descriptions of gravitational wave signals, as observed by a given detector $\mathcal{A}$. The random \emph{dots} $\left\{\theta_1,\cdots,\theta_N\right\}$ describe the intrinsic and extrinsic waveform properties. Such formalism allows to implement dots distributions and correlations with high degree of complexity (see~\cite{Buscicchio} for detailed derivations and~\cite{kampen1992stochastic} for a broader introduction to the topic).

For the sake of exposition, we restrict to the time domain and we isolate from $\theta_\sigma$ a parameter $\tau_\sigma$ associated with the random arrangement of the individual signals with respect to the $i$ index (e.g. time of arrivals). The remaining parameters will be referred to as $\hat{\theta}_\sigma$.
The statistical generative model reads as follows
\begin{align}
	N &\sim p ,\\
	\left\{\tau_\sigma\right\}_{1,\dots,N}\mid N &\sim Q_N ,\\
	\left\{\hat{\theta}_\sigma\right\}_{1,\cdots,N}\mid N &\sim P_N ,\\
	h^\mathcal{A}(t)  &=\sum_{\sigma=1}^{N}  u^\mathcal{A}(t-\tau_\sigma;\hat{\theta}_\sigma), \\
	\theta_\sigma &= (\tau_\sigma, \hat{\theta}_\sigma).
\end{align}

We will consider here a specific case of this model, known in literature as \emph{marked Campbell process}: independent identically distributed dots, characterized by a constant rate $\rho$ for the time domain and a single distribution for $\hat{\theta}_\sigma \sim p_\theta$.

For a realistic background, $\rho$ will be the total rate of all the events that contribute to the signal. The assumption of independent dots means that the events are not correlated, which is generally true for an astrophysical background on the time scale of the experiment if we neglect very peculiar scenarios, e.g. lensing effects. It should be noted that the formalism is flexible enough to be extended to such scenarios of correlated dots, by replacing $\rho$ with a more complex set of $Q_N, P_N$~\cite{kampen1992stochastic}.
The parameters $\hat{\theta}$ describe the event properties which we are interested in, e.g. their luminosity distance, their sky-position, the intrinsic source parameters. 
We will employ this machinery to evaluate the $h_i^\mathcal{A}$ cumulants
\begin{align}
	\Gamma^{\mathcal{A}_{1} \cdots \mathcal{A}_{n}}(t_1,\cdots,t_n) & = \langle \langle h^{\mathcal{A}_{1}}(t_1)\cdots h^{\mathcal{A}_{n}}(t_n)\rangle \rangle,
\end{align}
and we can replace such ensemble average, using $\langle u^{\cal A}\rangle_{\hat{\theta}} =0$, with
\begin{align}\label{eq:moment}
	\Gamma^{\mathcal{A}_{1} \cdots \mathcal{A}_{n}}(t_1,\cdots,t_n)\! =\rho\! \int\! \left\langle\prod_{k=1}^{n} u^{\mathcal{A}_{k}}(t-t_{k};\hat{\theta}) \right\rangle_{\!\hat{\theta}}\!dt .
\end{align}
The structure of this expression is quite straightforward to understand: 
contributions to the cumulants come only from the correlation of an event with itself, as seen by the chosen set of detectors. 
In principle, the procedure let us obtain a posterior probability distribution for the parameter's model, and upon suitable marginalization, for those of astrophysical interest: e.g., studying a background generated by coalescence events, the mass distribution as a function of redshift $z$.
Remarkably, correlations are not trivial as a consequence of the expectation value taken over the parameters, which makes them nonfactorized. Therefore each cumulant contains nontrivial and independent information about the parameter distributions, and it contributes directly to the inference in Eq.~\eqref{eq:statnongauss}. 
Moreover, it is worth highlighting an interesting scaling relation: scaling simultaneously the rate of events $\rho\rightarrow \rho^\prime=\alpha\rho$ and their amplitude $u\rightarrow u^\prime = \alpha^{-1/2} u$, cumulants of order $n$ become proportional to $\alpha^{1-n/2}$, i.e. for $n>2$ become negligible in the large $\rho$ limit while for $n=2$ they stay constant. This a simple manifestation of the central limit theorem.

Finally, we stress that our approach uses a population based construction of relevant cumulants: as a consequence, non-stationary noise contribution (i.e. glitches) can be absorbed in Eq.~\eqref{eq:camp} as additional population of signals~\cite{2022CQGra..39q5004A}--with different coupling to the detectors-- and integrated over in Eq.~\eqref{eq:moment}. This is subject of ongoing study.

\subsection{\label{subsec:Importance-sampling}Importance sampling}

The basic ingredient of the proposed approach is an efficient procedure to simulate a background with some target features. As we discussed in Sec.~\ref{subsec:Bayesian-approach} the building block is a procedure to generate a sample $h$ with the correct probability $p_h[h_{k+1}|{\cal M}_{k+1}]$ conditioned to a model ${\cal M}_{k+1}$.

A general parameterization of a given model can be given in terms of the event rate in a given volume of the parameter space, measured in the observer frame. This can be written as
\begin{equation}
	{\cal R}_0(\hat{\theta}) d\hat{\theta}_1 \cdots d\hat{\theta}_P .
\end{equation}
The total rate of events will be given by
\begin{equation}
	\rho = \int d\hat{\theta}_1 \cdots \int d\hat{\theta}_P {\cal R}_0(\hat{\theta})  
\end{equation}
and using it is possible to simulate dots in a given time interval. 
Notably, this rate can be very large, and it would be unfeasible to simulate in details all the events.
Instead, it is possible to introduce a threshold on events with negligible contribution to the background.
Alternatively, one can include it as a Gaussian contributions to the model. 
This is in fact one of the two reasons for introducing $g_i^{\cal A}$ in Eq.~\eqref{eq:decomposition}, the second being to include other Gaussian components, e.g. of cosmological origin.

Once the dots are generated, we ``decorate'' them by choosing a family of suitable individual waveforms and  associated  parameters according to their distribution $\rho^{-1} {\cal R}_0(\hat{\theta})$. Finally the strain $h_{ij}$ is generated, adding all contributions once projected onto each detector.

\section{Conclusions and perspectives}
\label{sec:Conclusions-and-perspectives}

In this paper, we propose a framework to construct detections statistics and perform Bayesian inference for non-Gaussian SGWBs.

The formalism is particularly suitable for stochastic backgrounds arising from the superposition of multiple overlapping sources. We discuss in details superposition in the time domain, but the approach can be generalized to the frequency domain.	 
We provide a recipe for computing the fundamental quantities required to perform our search in the realistic case of a SGWB of astrophysical origin. We do so by making use of marked Campbell processes.
We provide detailed derivations for a number of quantities related to the characterization of DSs performances, which we explore on a subset of representative points on the parameter space.

In a first application to a very simplified toy-model, comparatively to the standard approach to detection of Gaussian SGWBs, we observe significantly improved performances, in terms of the number of samples (i.e. the observation time or the frequency band) required to reach a target detection significance. 
As expected, this is milder in the presence of smaller non-Gaussianities.

Our approach is inherently complementary to those available in literature, since it rigorously models the SGWB as a stochastic signal, whose properties arise from the superposition of individual signals: we leverage the knowledge about their distribution and make use of a natural language suited to the purpose, i.e. marked Campbell processes.
We argue that the large flexibility attained in the data model through importance sampling motivates further studies on aspects crucial for a realistic application:
(i) backgrounds with non trivial overlap structure: a feature absent in our toy model, subject of ongoing study; 
(ii) superpositions of multiple backgrounds, as our framework offers a natural way to disentangle them; 
(iii) realistic noise models (non-stationary, correlated across detectors, non-Gaussian), to assess our approach performances compared to the ones in literature.

\begin{acknowledgments}

RB thanks E.~Buscicchio, F.~Di~Renzo, C.~J.~Moore and G.~Brocchi for useful conversations and stimulating observations.
RB acknowledges support through the Italian Space Agency grant \emph{Phase A activity for LISA mission, Agreement n. 2017-29-H.0, CUP F62F17000290005}.

\textit{Software}:
We acknowledge usage of 
\textit{Mathematica}~\cite{Mathematica}
and of the following
\textsc{Python}~\cite{10.5555/1593511}
packages for the analysis, post-processing and production of results throughout:
\textsc{CPNest}~\cite{2021zndo...4470001V},
\textsc{matplotlib}~\cite{2007CSE.....9...90H},
\textsc{numpy}~\cite{2020Natur.585..357H},
\textsc{scipy}~\cite{2020NatMe..17..261V}.

\end{acknowledgments}
	
\vfill

\bibliographystyle{apsrev4-2} 
\bibliography{bibliography}
\clearpage
\onecolumngrid
\appendix
\section{Detailed proofs}
\label{app:supplemental}
We expand here on definitions, assumptions, and more detailed derivations
for each expression in the paper, section by section. We will omit trivial
steps than can be performed easily with most symbolic computation
softwares. Moreover, we will omit full proofs, when a simplified version
contains already the interesting concepts. This is frequently the
case, e.g., for proofs given for single datapoint and/or single detector.

\subsection{Definitions \& Assumptions}

The signal is made of superposition of :
\begin{align}
	s_{i}^{\mathcal{A}} & =g_{i}^{\mathcal{A}}+h_{i}^{\mathcal{A}}+n_{i}^{\mathcal{A}}\label{eq:signal}.
\end{align}
The noise and the Gaussian background are distributed as 
\begin{align}
	p_{n}\left[n_{i}^{\mathcal{A}}\right] & =\mathcal{N}_{n}\exp\left(-\frac{1}{2}\mathcal{W}_{n}(n,n)\right) , \\
	p_{g}\left[g_{i}^{\mathcal{A}}\right] & =\mathcal{N}_{g}\exp\left(-\frac{1}{2}\mathcal{W}_{g}(g,g)\right) , \label{eq:pGaussian}
\end{align}
where the quadratic form $\mathcal{W}$ is defined by the sum of scalar products:

\begin{align}
	\mathcal{W}_{x}(u,v) & \equiv\sum_{\mathcal{A},\mathcal{B}}\mathcal{W}_{x}^{\mathcal{AB}}(u,v) ,\\
	\mathcal{W}_{x}^{\mathcal{AB}}(u,v) & =\left[\mathbb{C}_{xx}^{-1}\right]_{ij}^{\mathcal{AB}}u_{i}^{\mathcal{A}}v_{j}^{\mathcal{B}}.
\end{align}
The cross correlation array defines the normalization and the inner
structure of the quadratic form:
\begin{align}
	\mathcal{N}_{x} & =\exp\left(-\frac{1}{2}\text{Tr}\ln2\pi\mathbb{C}_{x}\right) ,\\
	\left[\mathbb{C}_{xy}\right]_{ij}^{\mathcal{AB}} & =\left\langle x_{i}^{\mathcal{A}}y_{j}^{\mathcal{B}}\right\rangle , \\
	\mathbb{C}_{x} & \equiv\mathbb{C}_{xx}.
\end{align}
The trace is performed over detector and data indices, and implicit summation over repeated indices is assumed. 
We make no assumptions on the distribution of $h$, $p_{h}[h_{i}^{\mathcal{A}}]$.

\subsection{The statistical problem}
We first prove Eq.~\eqref{eq:wiener}:
\begin{align}
	\mathcal{W}_{n+g}(u,v) & =\mathcal{W}_{n}(u,v)-\mathcal{G}(u,v).\label{eq:Gaussian_wiener}
\end{align}
This is a straightfoward application of the Woodbury identity: 
\begin{align}
	(A+UBV)^{-1} & =A^{-1}-A^{-1}U\left(B^{-1}+VA^{-1}U\right)^{-1}VA^{-1}\, ,
\end{align}
with $U=I$, $V=I$ 
and $A+B=\mathbb{C}_{n+g}=\mathbb{C}_{n}+\mathbb{C}_{g}$, $A=\mathbb{C}_{n}$. 
We therefore obtain
\begin{align}
	\mathbb{C}_{n+g}^{-1} & =\mathbb{C}_{n}^{-1}-\mathbb{C}_{n}^{-1}\left(\mathbb{C}_{g}^{-1}+\mathbb{C}_{n}^{-1}\right)^{-1}\mathbb{C}_{n}^{-1},
\end{align}
hence Eq.~(\ref{eq:wiener}).
Now we can prove Eq.~\eqref{eq:Gaussian_integral}
\begin{align}
	p_{s}[s] & =\mathcal{N}_{n+g}\int_{h}p_{h}[h]e^{-\frac{1}{2}\mathcal{W}_{n+g}(s-h,s-h)}.
\end{align}

The probability distribution of the data $s$ is specified by the knowledge of its components, and by the conditional probability
\begin{equation}\label{eq:conditional}
	p_{s}[s\mid h,g]=p_{n}[s-h-g]
\end{equation}
through Eq.~(\ref{eq:conditional}) we express Eq.~\eqref{eq:start} as 
\begin{align}
	p_{s}[s] & =\int_{h}\int_{g}p_{s}[s\mid h,g]p_{h}[h]p_{g}[g]\\
	& =\mathcal{N}_{n}\mathcal{N}_{g}\int_{h}\int_{g}p_{h}[h]e^{-\frac{1}{2}\left(\mathcal{W}_{n}(s-h-g,s-h-g)-\mathcal{W}_{g}(g,g)\right)}.\label{eq:doubleintegral}
\end{align}
The Gaussian integral on $g$ can be performed explicitly:
\begin{align}
	p_{s}[s]
	&=\int_{h}p_{h}[h]\int_g \mathcal{N}_n\mathcal{N}_g \exp
	\left[
	-\frac{1}{2}(s-h)^\top \mathbb{C}_{n}^{-1}(s-h)
	-\frac{1}{2}g^\top \mathbb{C}_{n}^{-1}g
	-\frac{1}{2}g^\top \mathbb{C}_{g}^{-1}g
	+(s-h)^\top \mathbb{C}_{n}^{-1}g
	\right]	
	\\
	&=
	\int_{h}p_{h}[h]\mathcal{N}_n\mathcal{N}_g\exp
	\left[
	-\frac{1}{2}(s-h)^\top \mathbb{C}_{n}^{-1}(s-h)
	\right]
	\int_g  \exp
	\left[
	-\frac{1}{2}
	g^\top
	\left[
	\mathbb{C}_{n}^{-1} +\mathbb{C}_{g}^{-1}
	\right]
	g
	+(s-h)^\top \mathbb{C}_{n}^{-1}g
	\right],
\end{align}
where ${}^\top$ denote transposing with respect to detectors and data indices.
Defining 
\begin{align}
	A&\equiv\mathbb{C}_{n}^{-1} +\mathbb{C}_{g}^{-1},\\
	v&\equiv A^{-1}\mathbb{C}_{n}^{-1}(s-h),
\end{align}
one gets:
\begin{align}
	\int_{h}p_{h}[h]\mathcal{N}_n\mathcal{N}_g\exp
	\left[
	-\frac{1}{2}(s-h)^\top \mathbb{C}_{n}^{-1}(s-h)
	\right]	
	\int_g  \exp
	\left[
	-\frac{1}{2}
	g^\top A g
	+v^\top A g
	\right]&=\\
	\int_{h}p_{h}[h]\mathcal{N}_n\mathcal{N}_g\exp
	\left[
	-\frac{1}{2}(s-h)^\top \mathbb{C}_{n}^{-1}(s-h)
	\right]	
	\int_g  \exp
	\left[
	-\frac{1}{2}(g-v)^\top A(g-v)
	+\frac{1}{2}v^\top A v
	\right].
\end{align} 
Integrating over $g$'s with fixed correlation matrix 
$\mathbb{C}_g$
\begin{align}
	\int_{h}p_{h}[h]\mathcal{N}_n\mathcal{N}_g\exp
	\left[
	-\frac{1}{2}(s-h)^\top
	\left[
	\mathbb{C}_{n}^{-1} -
	\mathbb{C}_{n}^{-1}
	(\mathbb{C}_{n}^{-1}+\mathbb{C}_{g}^{-1})^{-1}
	\mathbb{C}_{n}^{-1}
	\right]
	(s-h)
	\right]	
	\int_g  \exp
	\left[
	-\frac{1}{2}(g-v)^\top A(g-v)
	\right]
	&=\\
	\int_{h}p_{h}[h]\frac{
		2\pi^{-\frac{K}{2}}
		\sqrt{\det(\mathbb{C}_{n}^{-1})}
		\sqrt{\det(\mathbb{C}_{g}^{-1})}
	}
	{\sqrt{\det(\mathbb{C}_{n}^{-1}+\mathbb{C}_{g}^{-1})}}
	\exp
	\left[
	-\frac{1}{2}(s-h)^\top
	\left[
	\mathbb{C}_{n}^{-1} -
	\mathbb{C}_{n}^{-1}
	(\mathbb{C}_{n}^{-1}+\mathbb{C}_{g}^{-1})^{-1}
	\mathbb{C}_{n}^{-1}
	\right]
	(s-h)
	\right]
\end{align} 
with $K$ equal to the product between the number of detectors and the number of data points.
Using the Woodbury identity, Eq.~(\ref{eq:doubleintegral}) becomes
\begin{align}
	p_s\left[ s \right]
	&=
	\frac{2\pi^{-\frac{K}{2}}
	}
	{\sqrt{\det(\mathbb{C}_{g}(\mathbb{C}_{n}^{-1}+\mathbb{C}_{g}^{-1})\mathbb{C}_{n})}}
	\int_h 
	p_h\left[ h \right]
	\exp
	(
	-\frac{1}{2}(s-h)^\top
	\mathbb{C}_{n+g}^{-1}
	(s-h)
	)\\
	&=
	\frac{2\pi^{-\frac{K}{2}}
	}
	{\sqrt{\det\mathbb{C}_{n+g}}}
	\int_h 
	p_h\left[ h \right]
	\exp
	(
	-\frac{1}{2}(s-h)^\top
	\mathbb{C}_{n+g}^{-1}
	(s-h)
	).
\end{align} 
By interpreting the integral as an average over realizations
of $h$ distributed according to $p_{h}\left[\cdot\right]$ we obtain  Eq.~(\ref{eq:SignalProbability}).

\subsection{The Neyman-Pearson detection statistic}

We focus now on proving Eq.~\eqref{eq:Ystatsimple}.
From Eq.~\eqref{eq:Ystat} we obtain using assumptions from respective hypotheses
\begin{align}
	\log\frac{p_{s}\left[s\mid\mathcal{H}_{1}\right]}{p_{s}\left[s\mid\mathcal{H}_{0}\right]} & =\log\frac{\mathcal{N}_{n+g}}{\mathcal{N}_{n}}+\log\left\langle e^{-\frac{1}{2}\mathcal{W}_{n+g}(s,s)}e^{-\frac{1}{2}\mathcal{W}_{n+g}(h,h)}e^{\mathcal{W}_{n+g}(s,h)}\right\rangle _{\mathcal{H}_{1}}-\log\langle e^{-\frac{1}{2}\mathcal{W}_{n}(s,s)}\rangle_{\mathcal{H}_0}.
\end{align}
Averaging over $h$ at fixed data $s$ we obtain:
\begin{align}
	\log\frac{p_{s}\left[s\mid\mathcal{H}_{1}\right]}{p_{s}\left[s\mid\mathcal{H}_{0}\right]} & =\log\frac{\mathcal{N}_{n+g}}{\mathcal{N}_{n}}+\log\left\langle e^{-\frac{1}{2}\mathcal{W}_{n+g}(h,h)}e^{\mathcal{W}_{n+g}(s,h)}\right\rangle _{\mathcal{H}_{1}} -\frac{1}{2}\left(\mathcal{W}_{n+g}(s,s)-\mathcal{W}_{n}(s,s)\right)\\
	& =\log\frac{\mathcal{N}_{n+g}}{\mathcal{N}_{n}} + \log\left\langle e^{-\frac{1}{2}\mathcal{W}_{n+g}(h,h)}e^{\mathcal{W}_{n+g}(s,h)}\right\rangle _{\mathcal{H}_{1}}+\frac{1}{2}\mathcal{G}\left(s,s\right),\label{eq:general_detector}
\end{align}
hence Eq.~\eqref{eq:Ystatsimple}.

\subsubsection[Gaussian detection statistic]{Gaussian case}

The expansion of the DS reads as follows:
\begin{align}
	\hat{Y}(s)
	=\frac{1}{2} 
	\mathbb{C}_{n}^{-1}
	(\mathbb{C}_{n}^{-1} + \mathbb{C}_{g}^{-1})^{-1}
	\mathbb{C}_{n}^{-1}
	&=\frac{1}{2} 
	\mathbb{C}_{n}^{-1}
	(\mathbb{C}_{g}^{-1}
	(\mathbb{C}_{g}\mathbb{C}_{n}^{-1} + \mathbb{I})
	)^{-1}
	\mathbb{C}_{n}^{-1}\\
	&=\frac{1}{2} 
	\mathbb{C}_{n}^{-1}\mathbb{C}_{g}\mathbb{C}_{n}^{-1}
	+\mathcal{O}(\left\Vert\mathbb{C}_{g}\mathbb{C}_{n}^{-1}\right\Vert^2).
\end{align}
As in the main text, we start from the DS in Eq.~\eqref{eq:Ygauss}
\begin{align}
	\hat{Y}\left(s\right) 
	& 
	\simeq 
	\left[
	\frac{1}{2}
	\mathbb{\check{C}}_{n}^{-1}
	\check{\mathbb{C}}_{g}
	\check{\mathbb{C}}_{n}^{-1}
	\right]_{ij}^{\mathcal{AB}}
	s_{i}^{\mathcal{A}}
	s_{j}^{\mathcal{B}}
	=
	\check{\mathbb{A}}_{ij}^{\mathcal{AB}}s_{i}^{\mathcal{A}}s_{j}^{\mathcal{B}}.\label{eq:noise_bigger_signal}
\end{align}
We employ here estimates of $\mathbb{C}_{n}$,$\mathbb{C}_{g}$ labelled with a $\check{}$ symbol. 
They contain our prior knowledge about the noise and the Gaussian signal. 
The mean of
$\hat{Y}$ under $\mathcal{H}_{0}$ reads:

\begin{align}
	\label{eq:meanH0} 
	\mu_{\mathcal{H}_{0}} & =\frac{1}{2}\left[\check{\mathbb{C}}_{n}^{-1}\check{\mathbb{C}}_{g}\check{\mathbb{C}}_{n}^{-1}\right]_{ij}^{\mathcal{AB}}\langle{n_{i}^{\mathcal{A}}n_{j}^{\mathcal{B}}}\rangle 
	=\text{Tr}[\check{\mathbb{A}}\mathbb{C}_{n}].
\end{align}
Similarly for the variance (we drop the detector indices because they
follow the same contractions as the data indices)
\begin{align}
	\sigma_{\mathcal{H}_{0}}^{2} & =\check{\mathbb{A}}_{ij}^{\mathcal{AB}}\check{\mathbb{A}}_{kl}^{\mathcal{CD}}\langle{n_{i}^{\mathcal{A}}n_{j}^{\mathcal{B}}n_{k}^{\mathcal{C}}n_{l}^{\mathcal{D}}}\rangle-\left(\text{Tr}\left[\check{\mathbb{A}}\mathbb{C}_{n}\right]\right)^{2}\\
	& =\check{\mathbb{A}}_{ij}\check{\mathbb{A}}_{kl}\left(\left[\mathbb{C}_{n}\right]_{ik}\left[\mathbb{C}_{n}\right]_{jk}+\left[\mathbb{C}_{n}\right]_{il}\left[\mathbb{C}_{n}\right]_{jk}\right)\\
	& =\frac{1}{2}\text{Tr}\left[\check{\mathbb{C}}_{n}^{-1}\check{\mathbb{C}}_{g}\check{\mathbb{C}}_{n}^{-1}\mathbb{C}_{n}\check{\mathbb{C}}_{n}^{-1}\check{\mathbb{C}}_{g}\check{\mathbb{C}}_{n}^{-1}\mathbb{C}_{n}\right].\label{eq:varH0}
\end{align}
Similarly under $\mathcal{H}_{1}$ (using in addition $\langle{n_{i}^{\mathcal{A}}g_{j}^{\mathcal{B}}}\rangle=0$):
\begin{align}
	\mu_{\mathcal{H}_{1}} & =\frac{1}{2}\left[\check{\mathbb{C}}_{n}^{-1}\check{\mathbb{C}}_{g}\check{\mathbb{C}}_{n}^{-1}\right]_{ij}^{\mathcal{AB}}\langle{\left(n+g\right)_{i}^{\mathcal{A}}\left(n+g\right)_{j}^{\mathcal{B}}}\rangle\\
	& =\text{Tr}\left[\check{\mathbb{A}}\mathbb{C}_{n}\right]+\text{Tr}\left[\check{\mathbb{A}}\mathbb{C}_{g}\right]\\
	& =\mu_{\mathcal{H}_{0}}+\frac{1}{2}\text{Tr}\left[\check{\mathbb{C}}_{n}^{-1}\check{\mathbb{C}}_{g}\check{\mathbb{C}}_{n}^{-1}\mathbb{C}_{g}\right]\label{eq:meanH1supp}
\end{align}
and for the variance
\begin{align}
	\sigma_{\mathcal{H}_{1}}^{2} 
	& =\check{\mathbb{A}}_{ij}^{\mathcal{AB}}\check{\mathbb{A}}_{kl}^{\mathcal{CD}}\langle{\left(n+g\right)_{i}^{\mathcal{A}}\left(n+g\right)_{j}^{\mathcal{B}}\left(n+g\right)_{k}^{\mathcal{C}}\left(n+g\right)_{l}^{\mathcal{D}}}\rangle\\
	& =\check{\mathbb{A}}_{ij}^{\mathcal{AB}}\check{\mathbb{A}}_{kl}^{\mathcal{CD}}\left[\langle{n_{i}n_{j}n_{k}n_{l}}\rangle+\langle{n_{i}n_{j}g_{k}g_{l}}\rangle+\langle{n_{i}g_{j}g_{k}n_{l}}\rangle+\langle{n_{i}g_{j}n_{k}g_{l}}\rangle+\left(``n"\leftrightarrow``g"\right)\right] -\mu_{\mathcal{H}_{1}}^{2}.
\end{align}
All terms with an odd number of $n$'s cancel out after averaging
(both $g$ and $n$ are multivariate Gaussians). 
The fourth order averages simplify through Isserlis theorem to:
\begin{align}
	\sigma_{\mathcal{H}_{1}}^{2} 
	& =
	\check{\mathbb{A}}_{ij}\check{\mathbb{A}}_{kl}
	\left[
	\langle{n_{i}n_{j}}\rangle\,\langle{n_{k}n_{l}}\rangle
	+\langle{n_{i}n_{k}}\rangle\,\langle{n_{j}n_{l}}\rangle
	+\langle{n_{i}n_{l}}\rangle\,\langle{n_{k}n_{j}}\rangle
	\right]+\nonumber\\
	& +\check{\mathbb{A}}_{ij}\check{\mathbb{A}}_{kl}
	\left[
	\langle{n_{i}n_{j}}\rangle\,\langle{g_{k}g_{l}}\rangle
	\right]
	+\check{\mathbb{A}}_{ij}\check{\mathbb{A}}_{kl}
	\left[
	\langle{n_{i}n_{l}}\rangle\,\langle{g_{k}g_{j}}\rangle
	\right]+\nonumber\\
	& +\check{\mathbb{A}}_{ij}\check{\mathbb{A}}_{kl}
	\left[
	\langle{n_{i}n_{k}}\rangle\,\langle{g_{j}g_{l}}\rangle
	\right]
	+\left(``n"\leftrightarrow``g"\right)-\mu_{\mathcal{H}_{1}}^{2}.
\end{align}
Cancellations are again due to the uncorrelatedness and zero mean
of the two series. Upon contraction the expression simplifies to
\begin{align}
	\sigma_{\mathcal{H}_{1}}^{2} 
	& =\text{Tr}\left[\check{\mathbb{A}}\mathbb{C}_{n}\right]^{2}+2\text{Tr}\left[\check{\mathbb{A}}\mathbb{C}_{n}\check{\mathbb{A}}\mathbb{C}_{n}\right] +\text{Tr}\left[\check{\mathbb{A}}\mathbb{C}_{n}\right]\text{Tr}\left[\check{\mathbb{A}}\mathbb{C}_{g}\right]+2\text{Tr}\left[\check{\mathbb{A}}\mathbb{C}_{g}\check{\mathbb{A}}\mathbb{C}_{n}\right]+\left("n"\leftrightarrow"g"\right)-\mu_{\mathcal{H}_{1}}^{2}\\
	& =\sigma_{\mathcal{H}_{0}}^{2}+4\text{Tr}\left[\check{\mathbb{A}}\mathbb{C}_{n}\check{\mathbb{A}}\mathbb{C}_{g}\right]+\mathcal{O}\left(\left\Vert \mathbb{C}_{g}\right\Vert ^{2}\right) \simeq\sigma_{\mathcal{H}_{0}}^{2}+\text{Tr}\left[\check{\mathbb{C}}_{n}^{-1}\check{\mathbb{C}}_{g}\check{\mathbb{C}}_{n}^{-1}\mathbb{C}_{n}\check{\mathbb{C}}_{n}^{-1}\check{\mathbb{C}}_{g}\check{\mathbb{C}}_{n}^{-1}\mathbb{C}_{g}\right].\label{eq:sigmaH1}
\end{align}
Eqs.~\eqref{eq:meanH0},~\eqref{eq:varH0},~\eqref{eq:meanH1supp},
and (\ref{eq:sigmaH1}) prove results from the main text.

\subsubsection[Gaussian diagonal-free case]{Gaussian diagonal-free case}

Assuming uncorrelated noises across detectors (i.e. $\mathbb{\check{C}}_{n}^{\mathcal{AB}}\propto\delta^{\mathcal{AB}}$), we subtract by hand the diagonal terms from the statistics. We label the two ``diagonal-free'' hypotheses ${\mathcal{H}}_{0,\text{G}},{\mathcal{H}}_{1,\text{G}}$
and we have 
\begin{equation}
	\hat{Y}\left(s\right)=\frac{1}{2}\left[\mathbb{\check{C}}_{n}^{-1}\check{\mathbb{C}}_{g}\check{\mathbb{C}}_{n}^{-1}\right]_{ij}^{\mathcal{AB}}s_{i}^{\mathcal{A}}s_{j}^{\mathcal{B}}\longrightarrow\hat{Y}_G\left(s\right)=\frac{1}{2}\left[\mathbb{\check{C}}_{n}^{-1}\right]_{ik}^{\mathcal{AC}}\left[\check{\mathbb{C}}_{g}\right]_{kl}^{\mathcal{CD}}\left[\check{\mathbb{C}}_{n}^{-1}\right]_{lj}^{\mathcal{DB}}s_{i}^{\mathcal{A}}s_{j}^{\mathcal{B}}\left(1-\delta^{\mathcal{AB}}\right).
\end{equation}
Therefore we obtain a robust cross-correlation statistic (although not necessarily optimal) with the following properties:
\begin{align}
	\mu_{\mathcal{H}_{0,\text{G}}} 
	& \propto\frac{1}{2}\left[\mathbb{\check{C}}_{n}^{-1}\right]_{ik}^{\mathcal{AC}}\left[\check{\mathbb{C}}_{g}\right]_{kl}^{\mathcal{CD}}\left[\check{\mathbb{C}}_{n}^{-1}\right]_{lj}^{\mathcal{DB}}\delta^{\mathcal{AB}}\left(1-\delta^{\mathcal{AB}}\right)
	=0 \,,\\
	\mu_{\mathcal{H}_{1,\text{G}}} 
	& =\frac{1}{2}\left[\mathbb{\check{C}}_{n}^{-1}\right]_{ik}^{\mathcal{AC}}\left[\check{\mathbb{C}}_{g}\right]_{kl}^{\mathcal{CD}}\left[\check{\mathbb{C}}_{n}^{-1}\right]_{lj}^{\mathcal{DB}}\mathbb{C}_{g}^{\mathcal{AB}}\left(1-\delta^{\mathcal{AB}}\right)\\
	& =\frac{1}{2}\text{Tr}\left[\mathbb{\check{C}}_{n}^{-1}\check{\mathbb{C}}_{g}\check{\mathbb{C}}_{n}^{-1}\slashed{\mathbb{C}}_{g}\right],
\end{align}
where $\slashed{\mathbb{C}}_{g}^{\mathcal{AB}}=\mathbb{C}_{g}^{\mathcal{AB}}\left(1-\delta^{\mathcal{AB}}\right)$ defines a ``diagonal\textendash free'' signal correlation.
Since the noise is diagonal across detector indices, and detectors can have heterogeneous spectra, we write

\begin{align}
	\left[
	\mathbb{\check{C}}_{n}^{-1}
	\right]^{\mathcal{AC}} 
	& = 
	\sum_\epsilon
	c
	^{\mathcal{\epsilon}}
	\delta_{\epsilon}
	^{\mathcal{AC}}
	\\
	\delta_{\epsilon}^{\mathcal{AC}} & \equiv\begin{cases}
		1 & \epsilon=\mathcal{A}=\mathcal{C}\\
		0 & \text{otherwise}
	\end{cases}.
\end{align}
Then the DS becomes
\begin{align}
	\hat{Y}_G\left(s\right) 
	& =\frac{1}{2}c^{\mathcal{\epsilon}}_{ik}\delta_{\epsilon}^{\mathcal{AC}}\left[\check{\mathbb{C}}_{g}\right]_{kl}^{\mathcal{CD}}c_{lj}^{\mathcal{\delta}}\delta_{\delta}^{\mathcal{DB}}\left(1-\delta^{\mathcal{AB}}\right)s_{i}^{\mathcal{A}}s_{j}^{\mathcal{B}}.
\end{align}
Diagonal terms of $\check{\mathbb{C}}_{g}$ equals zero because
$\delta_{\epsilon}^{\mathcal{AC}}\delta_{\delta}^{\mathcal{DB}}\left(1-\delta^{\mathcal{AB}}\right)=0$
for $\mathcal{C=D}$. Therefore we are free to subtract them.
\begin{equation}
	\hat{Y}_G\left(s\right)=\frac{1}{2}c_{ik}^{\mathcal{\epsilon}}\delta_{\epsilon}^{\mathcal{AC}}\left[\check{\mathbb{C}}_{g}\right]_{kl}^{\mathcal{CD}}\left(1-\delta^{\mathcal{CD}}\right)c_{lj}^{\mathcal{\delta}}\delta_{\delta}^{\mathcal{DB}}\left(1-\delta^{\mathcal{AB}}\right)s_{i}^{\mathcal{A}}s_{j}^{\mathcal{B}}.
\end{equation}
For the same reason we can add them back in the rightmost factor,
i.e. neglecting $\left(1-\delta^{\mathcal{AB}}\right)$, because its
effect is now taken care of by $\delta^{\mathcal{CD}}$. In conclusion
\begin{align}
	\hat{Y}_G\left(s\right) & =\frac{1}{2}c_{ik}^{\mathcal{\epsilon}}\delta_{\epsilon}^{\mathcal{AC}}\left[\check{\mathbb{C}}_{g}\right]_{kl}^{\mathcal{CD}}\left(1-\delta^{\mathcal{CD}}\right)c_{lj}^{\mathcal{\delta}}\delta_{\delta}^{\mathcal{DB}}s_{i}^{\mathcal{A}}s_{j}^{\mathcal{B}}\\
	& =\frac{1}{2}\left[\mathbb{\check{C}}_{n}^{-1}\right]_{ik}^{\mathcal{AC}}\left[\check{\slashed{\mathbb{C}}}_{g}\right]_{kl}^{\mathcal{CD}}\left[\check{\mathbb{C}}_{n}^{-1}\right]_{lj}^{\mathcal{DB}}s_{i}^{\mathcal{A}}s_{j}^{\mathcal{B}}.
\end{align}
Therefore we can equivalently neglect the diagonal in our modelled signal
cross-correlation $\check{\mathbb{C}}_{g}$ or in the product of realizations
$s_{i}^{\mathcal{A}}s_{j}^{\mathcal{B}}$. Consequently, with obvious
definition
\begin{align}
	\check{\slashed{\mathbb{A}}}_{ij}^{\mathcal{AB}} & =\frac{1}{2}\left[\mathbb{\check{C}}_{n}^{-1}\check{\slashed{\mathbb{C}}}_{g}\check{\mathbb{C}}_{n}^{-1}\right]_{ij}^{\mathcal{AB}}
\end{align}
we obtain 
\begin{align}
	\mu_{\mathcal{H}_{0,\text{G}}} & 
	= 0 ,\\
	\mu_{\mathcal{H}_{1,\text{G}}} & =\text{Tr}\left[\check{\mathbb{A}}\slashed{\mathbb{C}}_{g}\right]=\text{Tr}\left[\check{\slashed{\mathbb{A}}}\mathbb{C}_{g}\right].
\end{align}

For the variances:
\begin{align}
	\sigma_{\mathcal{H}_{0,\text{G}}}^{2} & =\check{\slashed{\mathbb{A}}}_{ij}^{\mathcal{AB}}\check{\slashed{\mathbb{A}}}_{kl}^{\mathcal{CD}}
	\langle
	{
		n_{i}^{\mathcal{A}}
		n_{j}^{\mathcal{B}}
		n_{k}^{\mathcal{C}}
		n_{l}^{\mathcal{D}}
	}
	\rangle
	-\left(\text{Tr}\left[\check{\slashed{\mathbb{A}}}\mathbb{C}_{n}\right]\right)^{2}\\
	& =\frac{1}{4}\left[\mathbb{\check{C}}_{n}^{-1}\check{\slashed{\mathbb{C}}}_{g}\check{\mathbb{C}}_{n}^{-1}\right]_{ij}^{\mathcal{AB}}\left[\mathbb{\check{C}}_{n}^{-1}\check{\slashed{\mathbb{C}}}_{g}\check{\mathbb{C}}_{n}^{-1}\right]_{kl}^{\mathcal{CD}}\left(\left[\mathbb{C}_{n}\right]_{ik}^{\mathcal{AC}}\left[\mathbb{C}_{n}\right]_{jl}^{\mathcal{BD}}+\left[\mathbb{C}_{n}\right]_{il}^{\mathcal{AD}}\left[\mathbb{C}_{n}\right]_{jk}^{\mathcal{BC}}\right)\\
	& =2\text{Tr}\left[\check{\slashed{\mathbb{A}}}\mathbb{C}_{n}\check{\slashed{\mathbb{A}}}\mathbb{C}_{n}\right],\\
	\sigma_{\mathcal{H}_{1,\text{G}}}^{2} 
	& =\sigma_{\mathcal{H}_{0,\text{G}}}^{2}+\text{Tr}\left[\check{\mathbb{C}}_{n}^{-1}\check{\slashed{\mathbb{C}}}_{g}\check{\mathbb{C}}_{n}^{-1}\mathbb{C}_{n}\check{\mathbb{C}}_{n}^{-1}\check{\slashed{\mathbb{C}}}_{g}\check{\mathbb{C}}_{n}^{-1}\mathbb{C}_{g}\right]+\frac{1}{2}\text{Tr}\left[\check{\mathbb{C}}_{n}^{-1}\check{\slashed{\mathbb{C}}}_{g}\check{\mathbb{C}}_{n}^{-1}\mathbb{C}_{n}\check{\mathbb{C}}_{n}^{-1}\check{\slashed{\mathbb{C}}}_{g}\check{\mathbb{C}}_{n}^{-1}\mathbb{C}_{g}\right]\\
	& =\sigma_{\mathcal{H}_{0,\text{G}}}^{2}
	+4\text{Tr}\left[
	\check{\slashed{\mathbb{A}}}
	\mathbb{C}_{n}
	\check{\slashed{\mathbb{A}}}
	\mathbb{C}_{g}
	\right]+\mathcal{O}\left(\left\Vert \mathbb{C}_{g}\right\Vert ^{2}\right),
\end{align}
and hence Eqs~(\ref{eq:m0gauss}),(\ref{eq:m1gauss}),(\ref{eq:sigma0gauss}), and (\ref{eq:sigma1gauss}).
It is worth noting that $\sigma_{{\cal H}_{0}}, \sigma_{{\cal H}_{1}}$ and $\sigma_{{\cal H}_{0,\text{G}}}, \sigma_{{\cal H}_{1,\text{G}}}$ are respectively identical in functional form, with the substitution of $\mathbb{C}_g$ with its diagonal\textendash free version $\slashed{\mathbb{C}}_g$.

\subsubsection[Non-Gaussian case]{Non-Gaussian case}

We turn now our attention to the non-Gaussian case. By defining as
in the main text $\mathfrak{s}_{i}^{\mathcal{A}}=\left[\check{\mathbb{C}}_{n}^{-1}\right]_{ij}^{\mathcal{AB}}s_{j}^{\mathcal{B}}$,
direct substitution in the general expression of Eq.~\eqref{eq:Ystatsimple} yields
\begin{align}
	\hat{Y}\left(s\right) 
	&
	=\frac{1}{2}\mathcal{G}(s,s)+\log\left\langle e^{-\frac{1}{2}\mathcal{W}_{n+g}(h,h)}e^{\mathcal{W}_{n+g}(s,h)}\right\rangle \\
	& =\frac{1}{2}\left[\check{\mathbb{C}}_{g}\right]_{ij}^{\mathcal{AB}}\mathfrak{s}_{i}^{\mathcal{A}}\mathfrak{s}_{j}^{\mathcal{B}}+\log\chi_{h}\int_{h}p_{h}^{\prime}\left[h\right]e^{\mathcal{W}_{n+g}(s,h)},\label{eq:non-Gaussian-detector}
\end{align}
where we have defined a (normalized) probability distribution
\begin{align}
	p_{h}^{\prime}\left[h\right] & \equiv\chi_{h}^{-1}\exp\left[-\frac{1}{2}\mathcal{W}_{n+g}\left(h,h\right)\right]p_{h}\left[h\right]\\
	\chi_{h} & \equiv\left\langle e^{-\frac{1}{2}\mathcal{W}_{n+g}\left(h,h\right)}\right\rangle .
\end{align}
Now, focusing on 
\begin{equation}
	\log\chi_{h} \int_{h}p_{h}^{\prime}\left[h\right]e^{\mathcal{W}_{n+g}(s,h)},
\end{equation}
we expand $\mathcal{W}_{n+g}$ 
\begin{align}
	e^{\mathcal{W}_{n}\left(s,h\right)-\mathcal{G}\left(s,h\right)} & =\exp\left[{\mathfrak{s}_i^{\cal A}h_i^{\cal A}-\mathfrak{s}_{i}^{\mathcal{A}}\left[\check{\mathbb{C}}_{g}\right]_{ij}^{\mathcal{AB}}\left[\check{\mathbb{C}}_{n}^{-1}\right]_{jk}^{\mathcal{BC}}h_{k}^{\mathcal{C}}}\right].
\end{align}
We stress here again an important point: the separation of the signal
into a ``Gaussian'' and ``non-Gaussian'' component is somewhat
arbitrary. If we choose to set $g=0$, the entirety of the gravitational
wave signal is described by $h$. The double-whitened datapoints $\mathfrak{s}$
are not affected by this change, while Eq.~\eqref{eq:non-Gaussian-detector} becomes:

\begin{align}
	\hat{Y}\left(s\right) & =\log\left[\left\langle e^{-\frac{1}{2}\mathcal{W}_{n}\left(h,h\right)}\right\rangle \int_{h}p_{h}\left[h\right]e^{-\frac{1}{2}\mathcal{W}_{n}(h,h)}e^{\mathfrak{s}_i^{\cal A}h_i^{\cal A}}\right].
\end{align}
Therefore the rightmost term in Eq.~\eqref{eq:non-Gaussian-detector}
is the generating function of \emph{of the non-Gaussian component} ``connected moments'' (or ``cumulants''), with $\mathfrak{s}$ acting as the auxiliary variable, and realizations
$h$ distributed according to $p_{h}\left[h\right]e^{-\frac{1}{2}\mathcal{W}_{n}(h,h)}$,
\begin{align}
	\hat{Y}\left(s\right)=\log\left\langle e^{\mathfrak{s}_i^{\cal A}h_i^{\cal A}}\right\rangle .
\end{align}
Therefore by definition we can rewrite it as a power series in $\mathfrak{s}$,
\begin{align}
	\hat{Y}\left(s\right) & =\frac{1}{2}\left[\mathbb{C}_{g}\right]_{ij}^{\mathcal{AB}}\mathfrak{s}_{i}^{\mathcal{\mathcal{A}}}\mathfrak{s}_{j}^{\mathcal{\mathcal{B}}}+\chi_{h}\sum_{n=1}^{\infty}\frac{1}{n!}\check{\Gamma}_{i_{1}\dots i_{n}}^{\mathcal{A}_{1}\dots\mathcal{A}_{n}}\mathfrak{s}_{i_{1}}^{\mathcal{A}_{1}}\dots\mathfrak{s}_{i_{n}}^{\mathcal{A}_{n}}.
\end{align}
This proves the general expression for the non-Gaussian DS in Eq.~(\ref{eq:statnongauss}).
Evaluating the expectation values under both hypotheses, we get Eqs.(\ref{eq:mu0nongauss}) 
for
$\mathcal{H}_{0}$,
\begin{align}
	\mu_{\mathcal{H}_{0}} & =\text{Tr}\left[\check{\mathbb{A}}\mathbb{C}_{n}\right]+\chi_{h}\sum_{n=1}^{\infty}\frac{1}{n!}\check{\Gamma}_{i_{1}\dots i_{n}}^{\mathcal{A}_{1}\dots\mathcal{A}_{n}}\overline{\mathfrak{s}_{i_{1}}^{\mathcal{A}_{1}}\dots\mathfrak{s}_{i_{n}}^{\mathcal{A}_{n}}}\\
	& =\text{Tr}\left[\check{\mathbb{A}}\mathbb{C}_{n}\right]+\chi_{h}\sum_{n=1}^{\infty}\frac{1}{n!}\check{\Gamma}_{i_{1}\dots i_{n}}^{\mathcal{A}_{1}\dots\mathcal{A}_{n}}\mathbb{N}_{i_{1}\dots i_{n}}^{\mathcal{A}_{1}\dots\mathcal{A}_{n}}.
\end{align}
Using Isserlis theorem we express the $\mathbb{N}$'s as products of over
all pair of indices (i.e. over $\left[\mathbb{\check{C}}_{n}^{-1}\mathbb{C}_{n}\mathbb{\check{C}}_{n}^{-1}\right]_{ij}^{\mathcal{AB}}$).
Terms with odd $n$ cancel out to zero. Out of $n=2m$ indices, we
get $\left(2m-1\right)!!$ contractions in couples (the number of
complete graphs with $2m$ vertices), which due to the simmetry of
the $\Gamma$'s contribute identically after full contraction: 

\begin{align}
	\mu_{\mathcal{H}_{0}} & =\text{Tr}\left[\check{\mathbb{A}}\mathbb{C}_{n}\right]+\chi_{h}\underset{n=2m}{\sum_{m=1}^{\infty}}\frac{\left(n-1\right)!!}{n!}\check{\Gamma}_{i_{1}\dots i_{n}}^{\mathcal{A}_{1}\dots\mathcal{A}_{n}}\left[\mathbb{\check{C}}_{n}^{-1}\mathbb{C}_{n}\mathbb{\check{C}}_{n}^{-1}\right]_{i_{1}j_{2}}^{\mathcal{A}_{1}\mathcal{A}_{2}}\dots\left[\mathbb{\check{C}}_{n}^{-1}\mathbb{C}_{n}\mathbb{\check{C}}_{n}^{-1}\right]_{i_{n-1}j_{n}}^{\mathcal{A}_{n-1}\mathcal{A}_{n}}\\
	& =\text{Tr}\left[\check{\mathbb{A}}\mathbb{C}_{n}\right]+\chi_{h}\underset{n=2m}{\sum_{m=1}^{\infty}}\frac{1}{n!!}\check{\Gamma}_{i_{1}\dots i_{n}}^{\mathcal{A}_{1}\dots\mathcal{A}_{n}}\left[\mathbb{\check{C}}_{n}^{-1}\mathbb{C}_{n}\mathbb{\check{C}}_{n}^{-1}\right]_{i_{1}j_{2}}^{\mathcal{A}_{1}\mathcal{A}_{2}}\dots\left[\mathbb{\check{C}}_{n}^{-1}\mathbb{C}_{n}\mathbb{\check{C}}_{n}^{-1}\right]_{i_{n-1}j_{n}}^{\mathcal{A}_{n-1}\mathcal{A}_{n}},
\end{align}
and hence Eq.~(\ref{eq:mu0nongaussB}).
Results for the DS with ``scrambled-data'' [Eqs~(\ref{eq:m0nongaussianshift}) and \ref{eq:m1nongaussianshift}] are proven directly in the main text.

\subsection{Toy model derivations}\label{app:toy-model-derivation}
We show here the detailed derivations of the quantities related to the DS $\hat{Y}(s)$ for the toy model. We start from Eq.~\eqref{eq:Ystat}, and we drop the Gaussian component. Moreover we account for $s_{i},h_{i}$ being independent and diagonal across detectors, so we can factorize the $\left\langle \dots\right\rangle $ into product of expectation values for each $h_{i}$, obtaining
\begin{align}
	\hat{Y}\left(s\right) & =\log\left\langle \exp\left[-\frac{1}{2}\mathcal{W}_{n}(h,h)+\mathcal{W}_{n}(s,h)\right]\right\rangle \label{eq:tmYstatstart}\\
	& =\log\left\langle \exp\left[-\frac{1}{2}h_{i}^{\mathcal{A}}h_{j}^{\mathcal{B}}\left[\mathbb{C}_{n}^{-1}\right]_{\mathcal{AB}}^{ij}+h_{k}^{\mathcal{C}}s_{l}^{\mathcal{D}}\left[\mathbb{C}_{n}^{-1}\right]_{\mathcal{CD}}^{kl}\right]\right\rangle\\
	& =\log\left\langle \prod_{i}\exp\left[-\frac{1}{2}\frac{h_{i}^{\mathcal{A}}h_{j}^{\mathcal{B}}}{\sigma_{\mathcal{A}}^{2}}\delta_{\mathcal{AB}}\delta^{ij}+\frac{h_{k}^{\mathcal{C}}s_{l}^{\mathcal{D}}}{\sigma_{\mathcal{A}}^{2}}\delta_{\mathcal{CD}}\delta^{kl}\right]\right\rangle \\
	& =\sum_{i}\log\left\langle \exp\left[\sum_{\mathcal{A}}-\frac{h_{i}\left(h_{i}-2s_{i}^{\mathcal{A}}\right)}{2\sigma_{\mathcal{A}}^{2}}\right]\right\rangle , \label{eq:tmYstat}
\end{align}
where $\left\langle \dots\right\rangle $ in the last line is performed over a single datapoint $h_i$, but over multiple detectors. By using standard results on Gaussian integrals
and defining
\begin{align}
	A_\alpha&=\frac{1}{2\sigma_\alpha^2}+\sum_\mathcal{A}\frac{1}{2\sigma_\mathcal{A}^2},\\
	B_i&=\sum_\mathcal{A}\frac{s_i^\mathcal{A}}{\sigma_\mathcal{A}^2},
\end{align}
we obtain
\begin{align}
	\hat{Y}\left(s\right) 
	& = \sum_{i} \log\left[ \sum_{\alpha=+,-} \frac{\gamma_{\alpha}}{\sqrt{2\pi\sigma^2_{\alpha}}} 
	\int\mathrm{d}h_i\exp\left[-A_\alpha h_{i}^{2}+ B_ih_i\right]\right]
	=\sum_i\hat{y}\left(\sigma\sum_\mathcal{A}\frac{s_i^\mathcal{A}}{\sigma_\mathcal{A}^2}\right)
\end{align}
with
\begin{align}
	\hat{y}(u)&=\log \left[\sum_{\alpha=+,-} \frac{\gamma_{\alpha} \sigma}{\sqrt{\sigma_{\alpha}^{2}+\sigma^{2}}} \exp \left[\frac{\sigma_{\alpha}^{2}}{2\left(\sigma_{\alpha}^{2}+\sigma^{2}\right)} u^{2}\right]\right]\\
	\frac{1}{\sigma^{2}} & =\sum_{\mathcal{A}}\frac{1}{\sigma_{\mathcal{A}}^{2}},
\end{align}
and hence Eqs.~(\ref{eq:toyNPstatisticB}) to (\ref{eq:effective-noise}).
The basic building blocks for estimating the performances of the toy-model Neyman-Pearson DS are the means and variances of the $\hat{y}$ statistic.
It is a non-trivial function of a single scalar, combination of all the detector signals, under both hypotheses.
Evaluating them can be achieved numerically as follows. 

\subsubsection{Non-scrambled data}

Under the ${\cal H}_0$ hypothesis we need
\begin{align}
	\mu_0 & = \left< \hat{y}\left(u(s)  \right) \right> ,\\
	\sigma_0^2 & = \left< \hat{y}\left(u(s)   \right)^2 \right>-\mu_0^2,
\end{align}
where 
\begin{equation}
	u(s) = \sigma \sum_{\cal A} \frac{n_i^{\cal A}}{\sigma_{\cal A}^2}
\end{equation}
is a Gaussian variable with zero mean and unit variance.
Under ${\cal H}_1$ it $u(s)$ becomes
\begin{equation}
	u(s) = \sigma \sum_{\cal A} \frac{n_i^{\cal A}}{\sigma_{\cal A}^2}+  \frac{h_i}{\sigma} ,
\end{equation}
the sum of a normal variable (the first term) and a (scaled) variable distributed according to the mixture model. So the overall distribution is given by
\begin{equation}
	p_1(u) = \gamma_+ {\cal N}\left(u,\sqrt{1+\frac{\sigma_+^2}{\sigma^2}}\right)+\gamma_- {\cal N}\left(u,\sqrt{1+\frac{\sigma_-^2}{\sigma^2}}\right).
\end{equation}
Therefore we get 
\begin{align}
	\mu_0 & = \int du \hat{y}(u)\mathcal{N}(u,1),\\
	\sigma_0^2 & = \int du \hat{y}(u)^2 \mathcal{N}(u,1)-\mu_0^2 ,\\
	\mu_1 & = \int du \hat{y}(u) p_1(u) \label{eq:mu1_unshifted} ,\\
	\sigma_1^2 & = \int du \hat{y}(u)^2 p_1(u)-\mu_1^2 , \label{eq:sigma1_unshifted}
\end{align}
which can be easily evaluated numerically.

\subsubsection{Scrambled data}

The DS is
\begin{align}
	\mathring{Y}(s)
	&=\hat{Y}(s)-\hat{Y}(\mathring{s})\\
	&=\sum_i \hat{y}(u(s_i)) - \hat{y}(u(\mathring{s_i})) \label{eq:group}\\
	&=\sum_i \hat{y}(u(s_i)) - \sum_i\hat{y}(u(\mathring{s_i})).\label{eq:ungroup}
\end{align}
We need to evaluate the mean and the variance of the DS.
Under the hypothesis ${\cal H}_0$ we get
\begin{equation}
	u(\mathring{s_i})\mid \mathcal{H}_0 = \sigma \sum_{\cal A} \frac{\mathring{n}_i^{\cal A}}{\sigma_{\cal A}^2},
\end{equation}
and scrambling the noise realizations makes them uncorrelated across detectors; however the non-scrambled ones were already so.
So $u(\mathring{s}_i)$ is a zero-mean unit variance variable, and
therefore when evaluating the differences in \eqref{eq:group}, we have 
\begin{align}
	\mathring{\mu}_0 & = \left< \hat{y}\left( u(s)  \right) \right> - \left< \hat{y}\left( u(\mathring{s})  \right) \right>\,. \label{eq:mu0_shifted}
\end{align}
The two averages are identical, as $\mathring{s}$ and $s$ are identically distributed. Therefore $\mathring{\mu}_0=0$ as expected.
Computing explicitly the variance
\begin{align}
	\mathring{\sigma}_0^2
	&= 
	\left< 
	\left(
	\hat{y}(u(s))-\hat{y}(u(\mathring{s}))
	\right)^2 
	\right> -\mathring{\mu}_0^2 \\
	&=
	\left< 
	\hat{y}\left(u(s)\right)^2
	\right>
	+
	\left<
	\hat{y}\left(u(\mathring{s})\right)^2
	\right>
	-2
	\left<
	\hat{y}\left(u(s)\right)\right>\left<\hat{y}\left(u(\mathring{s})\right)
	\right>
	\\
	&=
	\sigma_0^2+\mu_0^2 +
	\sigma_0^2+\mu_0^2
	-2\mu_0^2
	\\
	&=2\sigma_0^2,
	\label{eq:missing}
\end{align}
where in Eq.~\eqref{eq:missing} we used the statistical independence by construction of the scrambled data $\mathring{s}$ upon the initial ones.
Under the hypothesis ${\cal H}_1$ we get
\begin{align}
	\mathring{\mu}_1 & = \left< \hat{y}\left( u(s)  \right) \right> - \left< \hat{y}\left( u(\mathring{s})  \right) \right>,
\end{align}
where now
\begin{equation}
	u(\mathring{s})= \sigma \sum_{\cal A} \frac{\mathring{n}_i^{\cal A}}{\sigma_{\cal A}^2} + \sigma \sum_{\cal A} \frac{\mathring{h}_i^{\cal A}}{\sigma_{\cal A}^2} .
\end{equation}
The first term is a normal Gaussian variable, as before. In the second term, each $x=\sigma\mathring{h}_i^{{\cal A}}/\sigma^2_{\mathcal{A}}$ is a different realization, distributed according to
\begin{equation}
	p_{\cal A}(x) = \gamma_+ {\cal N}\left(x,\frac{\sigma \sigma_+}{\sigma_{\cal A}^2}\right)+\gamma_- {\cal N}\left(x,\frac{\sigma \sigma_-}{\sigma_{\cal A}^2}\right),
\end{equation}
so the final distribution is given by the overall convolution
\begin{equation}
	\mathring{p}(u) = {\cal N}(u,1) \star p_1\left( u \right) \star \cdots \star  p_{N_D}\left( u \right)  ,
\end{equation}
where $N_D$ is the number of detectors available. This can be expressed as a sum of Gaussian distributions remembering that
\begin{equation}
	{\cal N}(u,\sigma_1) \star {\cal N}(u,\sigma_2) = {\cal N}\left(u,\sqrt{\sigma_1^2+\sigma_2^2}\right).
\end{equation}
In closed form, it reads
\begin{equation}
	\mathring{p}(u)=
	\sum_{k=0}^{N_D}
	\gamma_{+}^{k}\gamma_{-}^{N_D-k}
	\sum_{\mathtt{s}\in \mathcal{S}_{k}}
	\mathcal{N}\left(u, 
	\sqrt{
		1+
		\sigma^2\sum_{{\cal A}=1}^{N_D}\frac{\mathtt{s}_{\cal A}}{\sigma_{\cal A}^4}
	}
	\right),
\end{equation}
where $\mathcal{S}_{k}$ is the set of ordered sequences of $\sigma_{+}^2$  and $\sigma_{-}^2$, of length $N_D$, containing exactly $k$ $\sigma_{+}^2$s. For example, for $N_D=3$
\begin{align}
	\mathcal{S}_0&=\{\sigma_{-}^2,\sigma_{-}^2,\sigma_{-}^2\}, \\
	\mathcal{S}_1&=\{\sigma_{+}^2,\sigma_{-}^2,\sigma_{-}^2\},\{\sigma_{-}^2,\sigma_{+}^2,\sigma_{-}^2\},\{\sigma_{-}^2,\sigma_{-}^2,\sigma_{+}^2\}, \\
	\mathcal{S}_2&= \{\sigma_{+}^2,\sigma_{+}^2,\sigma_{-}^2\}, \{\sigma_{+}^2,\sigma_{-}^2,\sigma_{+}^2\},\{\sigma_{-}^2,\sigma_{+}^2,\sigma_{+}^2\}, \\
	\mathcal{S}_3&=\{\sigma_{+}^2,\sigma_{+}^2,\sigma_{+}^2\}.
\end{align}

So the new mean is corrected by a term $\mu_D$ with respect to the statistics on the non-scrambled data
\begin{align}
	\mathring{\mu}_1 & = \mu_1 - \mu_D \label{eq:mu1_shifted},\\
	\mu_D & = \int d\mathring{s}\hat{y}(u) \mathring{p}(u).
\end{align}
And similarly for the variance, which gets a correction $\sigma_D^2 = \int du \hat{y}(u)^2 \mathring{p}(u) - \mu_D^2$
\begin{align}
	\mathring{\sigma}_1^2 & = \left< ( \hat{y}\left( u(s)  \right)-\hat{y}\left( u(\mathring{s})  \right) )^2 \right> - \mathring{\mu}_1^2 \\
	& = \left< \hat{y}\left( u(s)  \right)^2 \right> + \left< \hat{y}\left( u(\mathring{s})  \right)^2 \right> - 2 \left< \hat{y}\left( u(s)  \right) \hat{y}\left( u(\mathring{s})  \right) \right> - \mathring{\mu}_1^2  \\
	& = \sigma_1^2+ \sigma_D^2 .\label{eq:sigma1_shifted}
\end{align}
In conclusion, with respect to Eq.~\eqref{eq:mu1_unshifted} and Eq.~\eqref{eq:sigma1_unshifted}, Eq.~\eqref{eq:mu1_shifted} and \eqref{eq:sigma1_shifted} constitute a correction to the DSs.

\subsubsection{Gaussian search of a non-Gaussian background}
If we ignore the non-Gaussianity of the toy model, and we model only its Gaussian part, we have
\begin{align}
	\left< s_i^{\cal A} s_j^{\cal B} \right> & = \delta^{\cal A B} \delta_{ij} \sigma_{\cal A}^2 \qquad \text{under } {\cal H}_0, \\
	\left< s_i^{\cal A} s_j^{\cal B} \right> & =   \delta_{ij} \left( \delta^{\cal A B} \sigma_{\cal A}^2 + \gamma_+ \sigma_+^2 + \gamma_- \sigma_-^2 \right) = \delta_{ij}\left(\delta^{\cal AB}\sigma^2_{\cal A} +\sigma_h^2\right) \text{under } {\cal H}_1.
\end{align}
The standard Gaussian detector is
\begin{align}
	\hat{Y}_G(s) = \sum_i \sum_{{\cal A}\neq{\cal B}} \frac{s_i^{\cal A} s_i^{\cal B}}{\sigma_{\cal A}^2 \sigma_{\cal B}^2}.
\end{align}

Without loss of generality, we focus on a single datapoint and omit the $i$ index. Under ${\cal H}_0$ we get mean and second order moment
\begin{align}
	\mu_{0,G}= \sum_{{\cal A}\neq{\cal B}} \frac{\left< s^{\cal A} s^{\cal B}\right>_0}{\sigma_{\cal A}^2 \sigma_{\cal B}^2} 
	= \sum_{{\cal A}\neq{\cal B}} \frac{\sigma_{\cal A}^2 \delta_{{\cal A B}} }{\sigma_{\cal A}^2 \sigma_{\cal B}^2} = 0
\end{align}
and
\begin{align}
	\sigma^2_{0,G} + \mu^2_{0,G} & = \sum_{{\cal A}\neq{\cal B}} \sum_{{\cal C}\neq{\cal D}} \frac{\left< s^{\cal A} s^{\cal B} s^{\cal C} s^{\cal D} \right>_0}{\sigma_{\cal A}^2 \sigma_{\cal B}^2 \sigma_{\cal C}^2 \sigma_{\cal D}^2}  \nonumber \\
	& = \sum_{{\cal A}\neq{\cal B}} \sum_{{\cal C}\neq{\cal D}} \frac{ \delta_{\cal A C} \delta_{\cal B D} \sigma_{\cal A}^2 \sigma_{\cal B}^2}{\sigma_{\cal A}^2 \sigma_{\cal B}^2 \sigma_{\cal C}^2 \sigma_{\cal D}^2} + \sum_{{\cal A}\neq{\cal B}} \sum_{{\cal C}\neq{\cal D}} \frac{ \delta_{\cal A D} \delta_{\cal B C} \sigma_{\cal A}^2 \sigma_{\cal B}^2}{\sigma_{\cal A}^2 \sigma_{\cal B}^2 \sigma_{\cal C}^2 \sigma_{\cal D}^2} \nonumber \\
	& = 2 \sum_{{\cal A}\neq{\cal B}} \frac{1}{\sigma_{\cal A}^2 \sigma_{\cal B}^2 } .
\end{align}

Under ${\cal H}_1$ we get
\begin{align}
	\mu_{1,G} = \sum_{{\cal A}\neq{\cal B}} \frac{\left< h^{\cal A} h^{\cal B}\right>_1}{\sigma_{\cal A}^2 \sigma_{\cal B}^2} 
	= \sigma_h^2 \sum_{{\cal A}\neq{\cal B}} \frac{1}{\sigma_{\cal A}^2 \sigma_{\cal B}^2} 
\end{align}
and
\begin{align}
	\sigma^2_{1,G} + \mu^2_{1,G} 
	& = \sum_{{\cal A}\neq{\cal B}} \sum_{{\cal C}\neq{\cal D}} \frac{\left< n^{\cal A} n^{\cal B} n^{\cal C} n^{\cal D} \right>_1}{\sigma_{\cal A}^2 \sigma_{\cal B}^2 \sigma_{\cal C}^2 \sigma_{\cal D}^2}
	+ \sum_{{\cal A}\neq{\cal B}} \sum_{{\cal C}\neq{\cal D}} \frac{\left< h^{\cal A} h^{\cal B} h^{\cal C} h^{\cal D} \right>_1}{\sigma_{\cal A}^2 \sigma_{\cal B}^2 \sigma_{\cal C}^2 \sigma_{\cal D}^2} \nonumber \\   
	&+ \sum_{{\cal A}\neq{\cal B}} \sum_{{\cal C}\neq{\cal D}} \frac{\left< n^{\cal A} h^{\cal B} n^{\cal C} h^{\cal D} \right>_1}{\sigma_{\cal A}^2 \sigma_{\cal B}^2 \sigma_{\cal C}^2 \sigma_{\cal D}^2}
	+ \frac{\left< n^{\cal A} h^{\cal B} h^{\cal C} n^{\cal D} \right>_1}{\sigma_{\cal A}^2 \sigma_{\cal B}^2 \sigma_{\cal C}^2 \sigma_{\cal D}^2}
	+ \frac{\left< h^{\cal A} n^{\cal B} n^{\cal C} h^{\cal D} \right>_1}{\sigma_{\cal A}^2 \sigma_{\cal B}^2 \sigma_{\cal C}^2 \sigma_{\cal D}^2}
	+ \frac{\left< h^{\cal A} n^{\cal B} h^{\cal C} n^{\cal D} \right>_1}{\sigma_{\cal A}^2 \sigma_{\cal B}^2 \sigma_{\cal C}^2 \sigma_{\cal D}^2} \\
	& = 2 \sum_{{\cal A}\neq{\cal B}}  \frac{1}{\sigma_{\cal A}^2 \sigma_{\cal B}^2} 
	+ 3 \left(\gamma_+ \sigma_+^4 + \gamma_- \sigma_-^4 \right) \left( \sum_{{\cal A}\neq{\cal B}}  \frac{1}{\sigma_{\cal A}^2 \sigma_{\cal B}^2} \right)^4
	+ 4 \sigma_h^2 \sum_{{\cal A}\neq{\cal B}} \sum_{{\cal A}\neq{\cal C}} \frac{1}{\sigma_{\cal A}^2 \sigma_{\cal B}^2 \sigma_{\cal C}^2}.
\end{align}	

\subsection{Bayesian analysis for the toy model}\label{app:posterior_toymodel}

The Bayesian inference can be constructed by parameterizing the signal hypothesis ${\cal H}_1$ with the model parameters $\cal M$. Therefore, the posterior reads: 
\begin{align}
	p\left(\mathcal{M}\mid s\right)=\mathcal{L}(s\mid {\cal M})\pi({\cal M})\propto\mathcal{N}_{n+g}\int_{h}e^{-\frac{1}{2}\mathcal{W}_{n+g}\left(s-h,s-h\right)}p_{h}\left[h\mid\mathcal{M}\right]\pi\left(\mathcal{M}\right).
\end{align}
For the toy model in Sec.~\ref{sec:toymodel}, $\mathcal{M}$ is specified by $\left(\sigma_{h},\sigma_{+},\sigma_{-}\right)$ and, assuming stationary, uncorrelated noises across detectors and $g=0$, the following simplifications occur:
\begin{align}
	\mathcal{N}_{n+g} = \mathcal{N}_{n}
	&
	= \left(\prod_{\mathcal{A}=1}^{N_D} \frac{1}{\sqrt{2\pi \sigma_{\cal A}^2}}\right)^{N_s},\\
	\mathcal{W}_{n+g}=\mathcal{W}_{n}\left(s-h,s-h\right)
	& = \sum_i \sum_{\cal A} \frac{\left(s^{\mathcal{A}}_i-h_{i}^{\mathcal{A}}\right)^2}{\sigma^2_{\cal A}}.
\end{align}
Therefore the posterior reads
\begin{align}
	p\left(\mathcal{M}\mid s\right)	
	\propto
	&\left(\prod_{{\cal A}=1}^{N_D}\frac{1}{\sqrt{2\pi }\sigma_{\cal A}}\right)^{N_s}\int \prod_i\mathrm{d}h_i\exp\left[-\frac{1}{2}\sum_i \sum_{\cal A} \frac{\left(s^{\mathcal{A}}_i-h_i\right)^2}{\sigma^2_{\cal A}} \right]p(h_i\mid {\cal M})\pi\left(\mathcal{M}\right),
\end{align}
which, as expected, decomposes into the product of likelihoods for each datapoint
\begin{align}
	p\left(\mathcal{M}\mid s\right)	&= \pi({\cal M})\prod_i\mathcal{L}(s_i\mid {\cal M}),\\	
	\mathcal{L}(s_i\mid{\cal M}) &=\left(\prod_{{\cal A}=1}^{N_D} \frac{1}{\sqrt{2\pi }\sigma_{\cal A}}\right) \int \mathrm{d}h\exp\left[-\frac{1}{2}\sum_{\cal A} \frac{\left(s^{\mathcal{A}}_i-h\right)^2}{\sigma^2_{\cal A}} \right]p(h\mid {\cal M}) \\
	&= 
	\sum_{\alpha=+,-} \frac{\gamma_\alpha}{\sqrt{1+\frac{\sigma^2_\alpha}{\sigma^2}}}
	\exp
	\left\{
	-\frac{1}{2} {\cal Q}_{\cal A B}^\alpha s_i^{\cal A} s_i^{\cal B}-\sum_{\cal A} \log \sqrt{2\pi} \sigma_{\cal A}
	\right\},
\end{align}
where the single datapoints collected across detectors are weighted by the quadratic form
\begin{align}
	{\cal Q}_{\cal A B}^\alpha & = \frac{1}{\sigma_{\cal A}^2 \sigma_{\cal B}^2} \left(\delta_{\cal A B}-\frac{\sigma_\alpha^2}{\sigma^2+\sigma_\alpha^2} \frac{\sigma_{\cal A}^{-1}\sigma_{\cal B}^{-1}}{\sigma^{-2}} \right).
\end{align}

\subsection{Cumulants for the toy model}\label{sec:cumulantstoymodel}
We provide here an explicit calculation for the cumulant generating function of the model presented in Sec.~\ref{sec:toymodel}. 
This serves the reader with a mapping of previous approaches in literature into our formalism~\cite{Martellini:2014xia}. It is straightforward to compute the cumulant generating function of a single $h$ distributed according to Eq.~\eqref{eq:prob-of-h}
\begin{equation}
	K(z) = \log\left[\gamma_{-}\exp\left({\frac{z^2\sigma_-^2}{2}}\right)+\gamma_+ \exp\left(\frac{z^2\sigma_+^2}{2}\right)\right] \,{.}
\end{equation}
Being each $h_i$ independent (upon scrambling) and equally distributed across detectors, the cumulant generating function of a set of $N$ datapoints is simply a sum of $K$'s with independent auxiliary variables.
\begin{equation}
	\label{eq:cgf-toymodel}
	\mathcal{K}\left(z_{i_{1}}^{\mathcal{A}_{1}},\dots,z_{i_{n}}^{\mathcal{A}_{n}}\right)=\sum_{j=1}^{n}K\left(z_{i_{j}}^{\mathcal{A}_{j}}\right).
\end{equation}
Previous studies approximate $p(h)$ with suitable asymptotic expansions (e.g. \emph{Gram-Charlier A} or \emph{Edgeworth} expansions), and then make use of a subset of cumulants. Those approach may be reproduced by using

\begin{align}
	\Gamma_{i_{1}\dots i_{n}}^{\mathcal{A}_{1}\dots\mathcal{A}_{n}} = & \,\partial_{i_{1}}^{\mathcal{A}_{1}}\dots\partial_{i_{n}}^{\mathcal{A}_{n}}\left.\mathcal{K}\left(z_{i_{1}}^{\mathcal{A}_{1}},\dots,z_{i_{n}}^{\mathcal{A}_{n}}\right)\right|_{z_{i_{j}}^{\mathcal{A}_{j}}=0}\\
	= & \,\mathbf{1}^{\mathcal{A}_{1}\dots\mathcal{A}_{n}}\delta_{i_{1}\dots i_{n}}n\left[\frac{\partial^{n}}{\partial s^{n}}K\left(s\right)\right]_{s=0}\\
	\equiv &  \,\mathbf{1}^{\mathcal{A}_{1}\dots\mathcal{A}_{n}}\delta_{i_{1}\dots i_{n}}n\Gamma_n .
\end{align}

Differentiating $k(s)$ yields the $n$-th cumulant $\Gamma_n$ of the single $h$. The value is, as a function of the parameter models $(\sigma_+,\sigma_-,\sigma_h)$ (assuming without loss of generality $\gamma_+>\gamma_-$)~\cite{Withers2015}

\begin{align}
	\Gamma_{2r} \,= & \,\delta_{1r}\sigma_{+}^{2}-\sum_{q=1}^{2r}\sum_{k=1}^{\infty}k^{q-1}\left(-\frac{\gamma_{-}}{\gamma_{+}}\right)^{k}B_{2r,q}\left(\sigma_{+}^{2},\sigma_{-}^{2}\right),\\
	\Gamma_{2r+1} \,= & \,0,
\end{align}
with $B_{k,q}$ the partial exponential Bell polynomials~\cite{Comtet:104089}.
\end{document}